\documentclass{emulateapj2}

\newcommand{\ldl}{$\lambda/{\Delta}{\lambda}$}
\newcommand{\logg}{$\log{g}$}
\newcommand{\teff}{T$_{eff}$}

\newcommand{\ammon}{NH$_3$}
\newcommand{\meth}{CH$_4$}
\newcommand{\wat}{H$_2$O}

\newcommand{\hh}{H$_2$}
\newcommand{\kms}{km~s$^{-1}$}
\newcommand{\cmss}{cm~s$^{-2}$}

\newcommand{\vtan}{$V_{tan}$}
\newcommand{\mjup}{M$_{\rm Jup}$}
\newcommand{\fsed}{$f_{sed}$}
\newcommand{\kzz}{$\kappa_{zz}$}
\newcommand{\logkzz}{$\log{\kappa_{zz}}$}

\slugcomment{Submitted 2011 March 11; Accepted to ApJ 2011 April 11}

\shorttitle{WISE T Dwarfs}
\shortauthors{Burgasser et al.}

\begin{document}

\title{FIRE Spectroscopy of Five Late-type T Dwarfs Discovered with the Wide-field Infrared Survey Explorer\footnote{This paper includes data gathered with the 6.5-m Magellan Telescopes located at Las Campanas Observatory, Chile.}}

\author{Adam J.\ Burgasser\altaffilmark{a,b,c},
Michael C.\ Cushing\altaffilmark{d},
J.\ Davy Kirkpatrick\altaffilmark{e},
Christopher R.\ Gelino\altaffilmark{e},
Roger L.\ Griffith\altaffilmark{e},
Dagny L.\ Looper\altaffilmark{f},
Christopher Tinney\altaffilmark{g},
Robert A.\ Simcoe\altaffilmark{b},
John J.\ Bochanski\altaffilmark{h},
Michael F.\ Skrutskie\altaffilmark{i},
A.\ Mainzer\altaffilmark{d},
Maggie A.\ Thompson\altaffilmark{j},
Kenneth A.\ Marsh\altaffilmark{e},
James M.\ Bauer\altaffilmark{d}, and
Edward L.\ Wright\altaffilmark{k}}

\altaffiltext{a}{Center for Astrophysics and Space Science, University of California San Diego, La Jolla, CA 92093, USA; aburgasser@ucsd.edu}
\altaffiltext{b}{Massachusetts Institute of Technology, Kavli Institute for Astrophysics and Space Research, 77 Massachusetts Avenue, Cambridge, MA 02139, USA}
\altaffiltext{c}{Hellman Fellow}
\altaffiltext{d}{NASA Jet Propulsion Laboratory, 4800 Oak Grove Drive,  Pasadena, CA 91109, USA}
\altaffiltext{e}{Infrared Processing and Analysis Center, MS 100-22, California Institute of Technology, Pasadena, CA 91125, USA}
\altaffiltext{f}{Institute for Astronomy, University of Hawaii, 2680 Woodlawn Drive, Honolulu, HI 96822, USA}
\altaffiltext{g}{Department of Astrophysics, School of Physics, University of New South Wales, NSW 2052, Australia}
\altaffiltext{h}{Department of Astronomy and Astrophysics, The Pennsylvania State University, University Park, PA 16802, USA}
\altaffiltext{i}{Department of Astronomy, University of Virginia,  Charlottesville, VA, 22904, USA}
\altaffiltext{j}{The Potomac School, 1301 Potomac School Road, McLean, VA 22101, USA}
\altaffiltext{k}{Department of Physics and Astronomy, UCLA, Los  Angeles, CA 90095-1562, USA}

\begin{abstract}
We present the discovery of five late-type T dwarfs identified with the Wide-field Infrared Survey Explorer (WISE).  Low-resolution near-infrared spectroscopy obtained with the Magellan Folded-port InfraRed Echellette (FIRE) reveal strong {\wat} and {\meth} absorption in all five sources, and spectral indices and comparison to spectral templates indicate classifications ranging from T5.5 to T8.5:. The spectrum of the latest-type source, WISE~J1812+2721,
is an excellent match to that of the T8.5 companion brown dwarf Wolf~940B.
WISE-based spectrophotometric distance estimates place these T dwarfs at 12-13~pc from the Sun,
assuming they are single.
Preliminary fits of the spectral data to the atmosphere models of Saumon \& Marley 
indicate effective temperatures ranging from 600~K to 930~K,
both cloudy and cloud-free atmospheres, and a broad range of ages and masses. 
In particular, two sources show evidence of both low surface gravity and cloudy atmospheres,
tentatively supporting a trend noted in other young brown dwarfs and exoplanets.
In contrast, the high proper motion T dwarf
WISE~J2018$-$7423 exhibits a suppressed $K$-band peak and
blue spectrophotometric $J-K$ colors indicative of an old, massive brown dwarf; however, it
lacks the broadened $Y$-band peak seen in metal-poor counterparts.  
These results illustrate the broad diversity of low-temperature brown dwarfs
that will be uncovered with WISE.
\end{abstract}

\keywords{
stars: fundamental parameters ---
stars: individual (WISEPC J161705.75+180714.0,
WISEPC J181210.85+272144.3,
WISEPC J201824.98$-$742326.1,
WISEPC J231336.41$-$803701.4,
WISEPC J235941.07$-$733504.8) ---
stars: low mass, brown dwarfs 
}

\section{Introduction}

The discovery in 1995 of a faint companion to the nearby M dwarf Gliese~229
galvanized the field of brown dwarf observational astrophysics.
Its near-infrared spectrum exhibits strong {\wat} and {\meth} absorption,
unambiguous indicators of a low-temperature,
substellar atmosphere \citep{1995Natur.378..463N, 1995Sci...270.1478O}.
These features now define the T dwarf spectral class \citep{2006ApJ...637.1067B},
the coldest known brown dwarfs with effective temperatures extending down to {\teff} $\approx$ 500~K (e.g., \citealt{2008MNRAS.391..320B, 2010MNRAS.408L..56L}).
Over 200 T dwarfs have been uncovered in the past 15 years,\footnote{For an up-to-date compilation, see the DwarfArchives website, \url{http://dwarfarchives.org}.} identified primarily 
in wide-field, near-infrared imaging surveys such as
the Two Micron All Sky Survey (2MASS; \citealt{2006AJ....131.1163S}; e.g., \citealt{2002ApJ...564..421B, 2007AJ....134.1162L}),
the Sloan Digital Sky Survey (SDSS; \citealt{2000AJ....120.1579Y}; e.g., \citealt{2002ApJ...564..466G, 2006AJ....131.2722C}),
the United Kingdom Infrared Telescope Deep Sky Survey (UKIDSS; \citealt{2007MNRAS.379.1599L}; e.g., \citealt{2007MNRAS.379.1423L, 2010MNRAS.406.1885B}),
and the Canada-France Brown Dwarf Survey (CFBDS; \citealt{2008A&A...484..469D, 2008A&A...482..961D, 2010A&A...522A.112R}).

Efforts are now underway to identify even colder brown dwarfs, sources whose 
atmospheres are anticipated to bridge the temperature gap between the known population and the Jovian planets ({\teff} $\lesssim$ 125~K).
This is the realm in which directly detectable young extrasolar planets are now being found
and investigated
(e.g., \citealt{2008Sci...322.1345K, 2008Sci...322.1348M, 2010ApJ...710L..35J, 2010ApJ...723..850B, 2010ApJ...721L.177C}).
Theoretical models of substellar atmospheres predict several interesting 
chemical transitions at these temperatures, including
the emergence of {\ammon} as a prominent absorber at near-infrared wavelengths;
the condensation of {\wat} and formation of thick ice clouds;
and the condensation of alkali salts, depleting brown dwarf atmospheres of 
spectrally prominent K~I and Na~I gases  \citep{1999ApJ...519..793L, 2002Icar..155..393L, 2002ApJ...568..335M, 2003ApJ...596..587B, 2006ApJ...647..552S, 2007ApJ...667..537L}.
All of these transitions have been suggested as possible triggers for
the definition of a new spectral class, tentatively designated the Y dwarf class \citep{2005ARA&A..43..195K}.  
These ``ultracold'' brown dwarfs may also comprise the bulk of the Galactic substellar population.  Depending on the underlying mass function, number densities for brown dwarfs colder than 600~K may exceed those of their warmer counterparts by a factor of a few or more \citep{2004ApJS..155..191B, 2008ApJ...689.1327S}. 
These sources would also sample the minimum brown dwarf formation mass, an important
statistic for brown dwarf formation theories (e.g., \citealt{2006A&A...458..817W})
and a determinant for the total baryonic mass associated with compact objects
(e.g., \citealt{1996ApJ...467L..65G}).

Two intriguing sources---both companions to nearby stars---have recently been identified whose estimated {\teff} $\approx$ 300--400~K
may broach the Y-dwarf regime \citep{2011ApJ...730L...9L, 2011arXiv1103.0014L}.  However, their
extremely faint near-infrared magnitudes ($J \gtrsim$ 21.5) 
have so far impeded spectroscopic follow-up.
In order to identify a significant sample of similarly cold brown dwarfs, search programs must shift to mid-infrared wavelengths where the majority of spectral flux emerges \citep{2003ApJ...596..587B}.   
The Wide-field Infrared Survey Explorer (WISE; \citealt{2010AJ....140.1868W}),
which has surveyed the full sky in four infrared bands centered at wavelengths of 3.4~$\micron$ ($W1$), 4.6~$\micron$ ($W2$), 12~$\micron$ ($W3$) and 22~$\micron$ ($W4$),  provides an opportunity to find these cold brown dwarfs.
The $W1$ and $W2$ bands were specifically designed to differentiate T dwarfs from background sources, sampling the strong 3.3~$\micron$ {\meth} band and the pseudo-continuum peak at 4.6~$\micron$, respectively
 \citep{1998ApJ...502..932O, 2003ApJ...596..587B, 2004AJ....127.3516G}.
\citet{2011ApJ...726...30M} have recently reported the first cold brown dwarf discovery with WISE, WISEPC~J045853.90+643451.9 (hereafter WISE~J0458+6434), 
a source which exhibits nearly saturated near-infrared {\wat} and {\meth} bands consist with a $\sim$T9 classification (see also \citealt{gelino2011}).

In this article, we report the discovery of five new late-type T dwarfs identified in WISE and confirmed through near-infrared spectroscopy with the 
Folded-port Infrared Echellette (FIRE;  \citealt{2008SPIE.7014E..27S,2010SPIE.7735E..38S}).
In Section~2 we describe the selection of these sources based on WISE photometry
and additional survey data. 
In Section~3 we describe our follow-up imaging and spectroscopic observations
that confirm the T dwarf nature of these sources.
In Section~4 we derive spectral classifications using both spectral indices and comparison to near-infrared spectral templates, and estimate distances and kinematics.
In Section~5 we provide additional constraints on the atmospheric and physical properties of these sources through spectral model fits using the calculations of \citet{2008ApJ...689.1327S}.
In Section~6 we discuss the properties of individual discoveries in detail.
Results are summarized in Section~7.

\section{Candidate Selection}

\subsection{WISE Photometry}

\begin{figure*}[t]
\centering
\includegraphics[width=0.7\textwidth,angle=90]{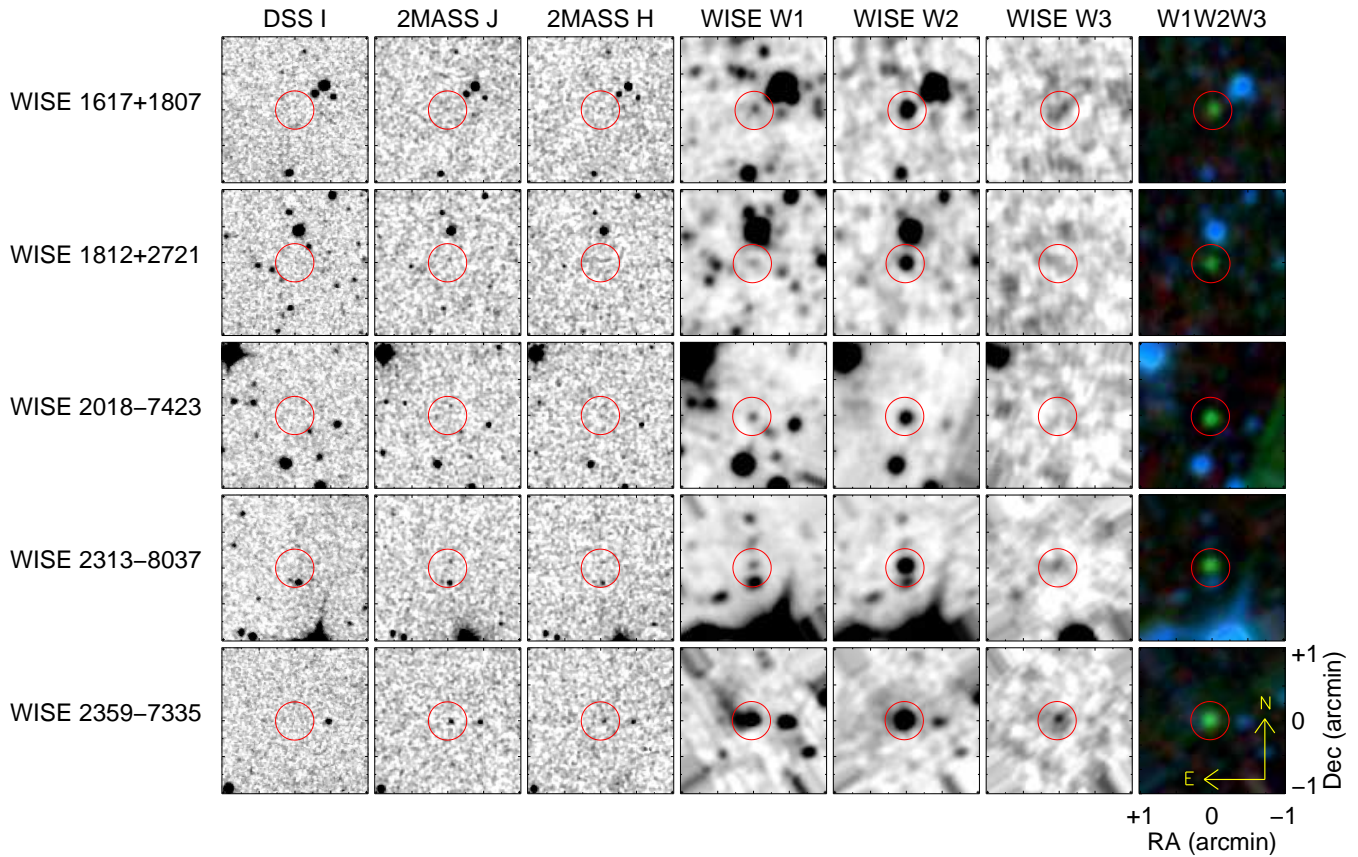}
\caption{Finderchart images of the five WISE T dwarfs, showing 2$\arcmin$$\times$2$\arcmin$ fields oriented with North up and East toward the left.  DSS $I_N$, 2MASS $JH$ and WISE $W1W2W3$ images are centered on the WISE coordinates for each source (red circle).  The rightmost image shows a false color composite of the three WISE images, with blue, green and red represented by $W1$, $W2$ and $W3$, respectively.
\label{fig_finders}}
\end{figure*}

Candidate late-type T dwarfs were selected from the WISE coadd source working database, as described in detail in \citet{2011ApJ...726...30M} and J.\ D.\ Kirkpatrick et al.\ (in prep.).   In brief, sources were selected to have $W1-W2$ $\geq$ 2, $W2-W3$ $\leq$ 2.5 (to exclude extragalactic sources; see \citealt{2010AJ....140.1868W}), a $W2$ signal-to-noise ratio $\geq$10, and a point spread function consistent with an unresolved point source.  These sources were then compared to optical and near-infrared imaging survey data from the Digitized Sky Survey (DSS), SDSS and 2MASS to exclude optical counterparts and other contaminants.  
The five new T dwarfs presented here represent only a subset of the full candidate pool currently under investigation.  Their designations\footnote{Throughout the text, we use the shorthand notation WISE~Jhhmm$\pm$ddmm to refer to WISE sources, where the suffix is the sexagesimal Right Ascension (hour and minute) and declination (degree and arcminute).} and measured photometry (excluding $W4$) are listed in Table~\ref{tab_sources}. Figure~\ref{fig_finders} displays DSS, 2MASS and WISE images of the fields around each target.

\begin{deluxetable*}{lcccccccccc}
\tabletypesize{\tiny}
\tablewidth{0pt}
\tablecaption{Photometric Data for WISE Brown Dwarf Discoveries\label{tab_sources}}
\tablehead{
\colhead{Designation} &
\colhead{$Y$} &
\colhead{$J$} &
\colhead{$H$} &
\colhead{$K_s$} &
\colhead{$W1$} &
\colhead{$W2$} &
\colhead{$W3$} &
\colhead{$H-W2$} &
\colhead{$W1-W2$} &
\colhead{$W2-W3$} \\
 &
\colhead{(mag)} &
\colhead{(mag)} &
\colhead{(mag)} &
\colhead{(mag)} &
\colhead{(mag)} &
\colhead{(mag)} &
\colhead{(mag)} &
\colhead{(mag)} &
\colhead{(mag)} \\
}
\startdata
WISEPC~J161705.75+180714.0  & 18.71$\pm$0.04 & 17.66$\pm$0.08 & 18.23$\pm$0.08 & \nodata & 16.89$\pm$0.16 &  14.03$\pm$0.06 &   12.48$\pm$0.48 & 4.20$\pm$0.10 & 2.86$\pm$0.17 &  1.55$\pm$0.48       \\
WISEPC~J181210.85+272144.3  & \nodata & 18.19$\pm$0.06 & 18.83$\pm$0.16 & \nodata & 17.32$\pm$0.21 &    14.15$\pm$0.05 & $>$12.4 &      4.68$\pm$0.17    & 3.17$\pm$0.21 &  $<$1.8         \\
WISEPC~J201824.98$-$742326.1 & \nodata & 17.11$\pm$0.21\tablenotemark{a} & $>$16.5\tablenotemark{a} & $>$16.6\tablenotemark{a}  & 16.55$\pm$0.10 &   13.76$\pm$0.03 &   $>$12.3        &     $>$2.7     & 2.80$\pm$0.11 &  $<$1.5         \\
WISEPC~J231336.41$-$803701.4 &  \nodata & 16.97$\pm$0.24\tablenotemark{a} & $>$16.2\tablenotemark{a} & $>$16.4\tablenotemark{a} &  16.29$\pm$0.07 &   13.77$\pm$0.04 &   12.52$\pm$0.32 &     $>$2.4        &   2.52$\pm$0.08 &  1.25$\pm$0.33  \\
WISEPC~J235941.07$-$733504.8 & \nodata & 16.17$\pm$0.04\tablenotemark{a} & 16.07$\pm$0.07\tablenotemark{a} & 16.05$\pm$0.13\tablenotemark{a} & 15.12$\pm$0.04 & 13.26$\pm$0.04 & 11.63$\pm$0.20 & 2.65$\pm$0.19 & 1.86$\pm$0.06 & 1.63$\pm$0.20 \\
\enddata
\tablenotetext{a}{Photometry from the 2MASS Reject Table (WISE~J2018$-$7423 and WISE~J2313$-$8037) and 6x Catalogs (WISE~J2359$-$7335; \citealt{2006AJ....131.1163S}).} 
\end{deluxetable*}

\subsection{Additional Survey Photometry and Astrometry}

All of the WISE targets were cross-matched to the 2MASS, SDSS and UKIDSS catalogs.
One source, WISE~J2359$-$7335, had a counterpart in the 2MASS Point Source and 6x catalogs\footnote{This source had been previously identified by D.\ Looper as part of a color-selected search of the 2MASS 6x catalog, but not published; see \citet{2007AJ....134.1162L}. It is included here despite having $W1-W2 < 2$.},
with a $J$ magnitude in the latter of 16.17$\pm$0.04.  
WISE~J2018$-$7423 and WISE~J2313$-$8037 had faint counterparts in the 2MASS Reject Catalog, with $J$ = 17.11$\pm$0.21 and 16.97$\pm$0.24~mag, respectively (detection grades of ``C'' and ``E''), located $\sim$7-10$\arcsec$ from their WISE positions. 
We confirmed these counterparts were associated using our $J$-band FIRE acquisition images (see Section~3.2.1).
WISE~J1617+1807 has a nearby counterpart in the SDSS Data Release 7 catalog 9$\arcsec$ from its WISE position, but this match appears spurious based on the source's blue optical colors ($i$ = 21.91$\pm$0.16, $i-z < -0.6$).
No common proper motion companions were found within 5$\arcmin$ of any of the WISE sources in SIMBAD or in the US Naval Observatory CCD Astrograph Catalog (UCAC3; \citealt{2010AJ....139.2184Z}).

\begin{deluxetable*}{lllcccc}
\tabletypesize{\scriptsize}
\tablewidth{0pt}
\tablecaption{Proper Motions for WISE T Dwarfs Detected in 2MASS\label{tab_astrometry}}
\tablehead{
 & \multicolumn{2}{c}{Astrometry} &
\colhead{$\Delta{t}$} &
\colhead{$\mu_{\alpha}\cos{\delta}$} &
\colhead{$\mu_{\delta}$} &
\colhead{\vtan\tablenotemark{a}}  \\
\cline{2-3}
\colhead{Source} &
\colhead{2MASS} &
\colhead{WISE} &
\colhead{(yr)} &
\colhead{(mas~yr$^{-1}$)} &
\colhead{(mas~yr$^{-1}$)} &
\colhead{({\kms})}  \\
}
\startdata
WISE~J2018$-$7423 & 20:18:24.24 $-$74:23:17.92 & 20:18:24.98 $-$74:23:26.14 & 9.6 & 311$\pm$32 & $-$852$\pm$31 & 56$\pm$6 \\
WISE~J2313$-$8037 & 23:13:35.37 $-$80:36:56.24  & 23:13:36.40 $-$80:37:01.40 & 10.4 & 242$\pm$32 & $-$496$\pm$29 & 31$\pm$5 \\
WISE~J2359$-$7335 & 23:59:40.33 $-$73:35:05.33 & 23:59:41.07 $-$73:35:04.87 & 9.5 & 332$\pm$26 & 49$\pm$23 & 20$\pm$3 \\
\enddata
\tablenotetext{a}{Based on the spectrophotometric distance estimates listed in Table~\ref{tab_characteristics}.} 
\end{deluxetable*}

The roughly ten-year baseline between the 2MASS and WISE detections of WISE~J2018$-$7423, WISE~J2313$-$8037 and WISE~J2359$-$7335 allows proper motion measurements for these sources.  Astrometry from the two catalogs are listed in Table~\ref{tab_astrometry},
and the computed proper motions incorporate uncertainties in the 2MASS and WISE positions but do not account for parallactic motion.  
We note that an error in the astrometric calibration pipeline of the WISE working database (now corrected) leads to an occasional large offset (of order 1$\arcsec$) in declination coordinate.  As such, these proper motions should be considered preliminary until the WISE Final Release catalog astrometry is available.  Nevertheless, the relatively large angular
motions of these sources, as high as 0$\farcs$91$\pm$0$\farcs$03~yr$^{-1}$ for WISE~J2018$-$7423,  are typical for nearby field dwarfs.  

\section{Observations}

\subsection{Imaging}

\subsubsection{SOAR/SpartanIRC}

$JH$-band photometry of WISE~J1617+1807 were obtained on 2010 March 21 (UT) in clear conditions with the Spartan Infrared Camera (SpartanIRC; \citealt{2004SPIE.5492.1644L}) on the 4.1m SOAR telescope. 
The source was observed at an airmass of 1.53 with five 60-s exposures in each filter, dithered in   40$\arcsec$ offsets.
Imaging data were reduced using custom routines that perform 
flat-fielding and sky-subtraction, the latter from a sky frame 
created from a median stack of the dither image sequence.  
A 2$\arcmin$$\times$2$\arcmin$ mosaic was created by stacking the reduced images to a
common center and averaging.  Aperture photometry was measured for 
all sources in the mosaic, and photometric calibration was done on the 2MASS system
using three bright stars in the field-of-view, with a zero-point uncertainty estimated from
the standard deviation of photometric offsets for these three stars. 
Measurements are listed in Table~\ref{tab_sources}.

\subsubsection{Fan Mountain/FanCam}
           
$Y$-band photometry of WISE~J1617+1807 was obtained on 2010 April 1 (UT) in photometric conditions with the FanCam near-infrared imager mounted on the 0.8~m Fan Mountain telescope \citep{2009PASP..121..885K}. Imaging data were obtained and reduced as described in \citet{2011ApJ...726...30M}, with 15 exposures of 60~s and 80 exposures of 30~s obtained in a 15$\arcsec$ dither pattern, for a total exposure time of 55 minutes.  Aperture photometry on the mosaicked frame was measured using standard IRAF routines, with an instrumental zeropoint derived by estimating $Y$-band magnitudes of neighboring stars from their 2MASS $J$ and $K_s$ photometry and the transformation of \citet{2006PASP..118....2H}.  We found $Y$ = 18.71$\pm$0.04 for WISE~J1617+1807, implying $Y-W2$ =  4.68$\pm$0.07, about 0.7~mag bluer than WISE~J0458+6434 \citep{2011ApJ...726...30M}.

\subsubsection{Palomar/WIRC}

$JH$-band photometry of WISE~J1812+2721 were obtained on 2010 Aug 30 (UT) using the
Wide-Field Infrared Camera (WIRC; \citealt{2003SPIE.4841..451W}) mounted on the 200-inch Hale
Telescope at Palomar Observatory.  WIRC has a pixel scale of
0$\farcs$2487~pixel$^{-1}$ and a total field of view of 8$\farcm$7$\times$8$\farcm$7.  
Conditions were clear during the observations, but with high humidity and poor seeing 
($\sim$2.5$\arcsec$ at $J$).  For each filter, a series of 15 exposures of 60~s each were obtained, 
dithering by 50-100$\arcsec$. Targets were observed over an airmass range of 1.09--1.13.

The imaging data were reduced using a suite of IRAF\footnote{Image Reduction and Analysis Facility (IRAF; \citealt{1986SPIE..627..733T}).} and FORTRAN programs 
provided by T.\ Jarrett.  These routines first linearize and dark subtract the
images, then create a sky frame and flat field images for 
each dither set which are subtracted from and divided into (respectively) 
each science image.
At this stage, WIRC images still contain a significant bias that is not
removed by the flat field.  Comparison of 2MASS and WIRC photometric
differences across the array shows that this flux bias has a level of
$\approx$10\% and the pattern is roughly the same for all filters.  Using
these 2MASS-WIRC differences for many fields, we created a flux bias
correction image that was applied to each of the ``reduced'' images.
Finally, we determined an astrometric calibration for the images using 2MASS stars in the
field, and the images were mosaicked together.  This final mosaic was
photometrically calibrated using 2MASS stars, and magnitudes computed
using aperture photometry.
Measurements are listed in Table~\ref{tab_sources}.

\subsection{Spectroscopy}

\subsubsection{Magellan/LDSS-3}

Optical spectroscopy of WISE~J2359$-$7335 was obtained on 2005 December 2 (UT) in clear conditions using the Low Dispersion Survey Spectrograph (LDSS-3; \citealt{1994PASP..106..983A})
mounted on the Magellan 6.5m Clay Telescope (see Table~\ref{tab_specobs} for a complete observing log).
Data were obtained using the VPH-red grism (660 lines/mm) and the 0$\farcs$75
(4-pixel) wide longslit, aligned along the parallactic angle, providing 6050--10500~{\AA}
spectroscopy with {\ldl} $\approx$ 1800.  The OG590 longpass filter
was used to eliminate second order light shortward of 6000 {\AA}.
Two exposures of 1800~s each were obtained over an airmass range of 1.62--1.81.
This was followed by observation of the nearby G2~V star HD~10991 ($V$ = 9.38) 
for telluric absorption correction. HeNeAr arc lamp and flat-field
quartz lamp exposures reflected off of the Clay secondary flat field screen
were obtained for dispersion and pixel
response calibration.  The data were reduced using the IRAF {\it onedspec} package,
as described in \citet{2007ApJ...657..494B}.

\begin{deluxetable*}{lllccll}
\tabletypesize{\scriptsize}
\tablewidth{0pt}
\tablecaption{Spectroscopic Observations\label{tab_specobs}}
\tablehead{
\colhead{Source} &
\colhead{Instrument} &
\colhead{UT Date} &
\colhead{Integration} &
\colhead{Airmass} &
\colhead{Calibrator}  &
\colhead{Conditions/}  \\
 &  & &
\colhead{(s)} & &
\colhead{Star} &
\colhead{Seeing} \\
}
\startdata
WISE~J1617+1807  & Magellan/FIRE & 2010 Apr 7 &   282 &  1.74   & BD+29~3523 & clear, 0$\farcs$6 \\
WISE~J1812+2721  & Magellan/FIRE & 2010 Sep 19 &   1045 &  1.88-2.03   & BD+30~3488   & clear, 1$\farcs$5--2$\arcsec$   \\
WISE~J2018$-$7423 & Magellan/FIRE & 2010 Sep 20 &   526 &  1.53-1.55   & HD~189588  &  clear, 0$\farcs$4  \\
WISE~J2313$-$8037 &  Magellan/FIRE & 2010 Sep 20 &   526 &  1.61   & HD~189588 & clear, 0$\farcs$4  \\
WISE J2359$-$7335 &  Magellan/LDSS-3 & 2005 Dec 2 &   3600 &  1.62-1.81   & HD~10991 & clear, 0$\farcs$8  \\
&  AAT/IRIS2 & 2006 May 15 &   600/600\tablenotemark{a} &  1.49/1.60\tablenotemark{a}   & HIP~118079 & clear \& humid, 1$\farcs$5  \\
&  AAT/IRIS2 & 2006 June 11 &   600 &  1.52   & HIP~118079 & clear \& humid, 2$\arcsec$  \\
&  Magellan/FIRE & 2010 Dec 24 &   846 &  2.02-2.14   & HD~189588 & clear, 0$\farcs$8  \\
Wolf 940B & Magellan/FIRE & 2010 Sep 19 &   1045 &  1.14-1.15   & HD~208368 & clear, 1$\farcs$5--2$\arcsec$ \\
\enddata
\tablenotetext{a}{For $J_l$ and $H_s$ observations, respectively.}
\end{deluxetable*}

\begin{figure*}
\centering
\epsscale{0.9}
\plotone{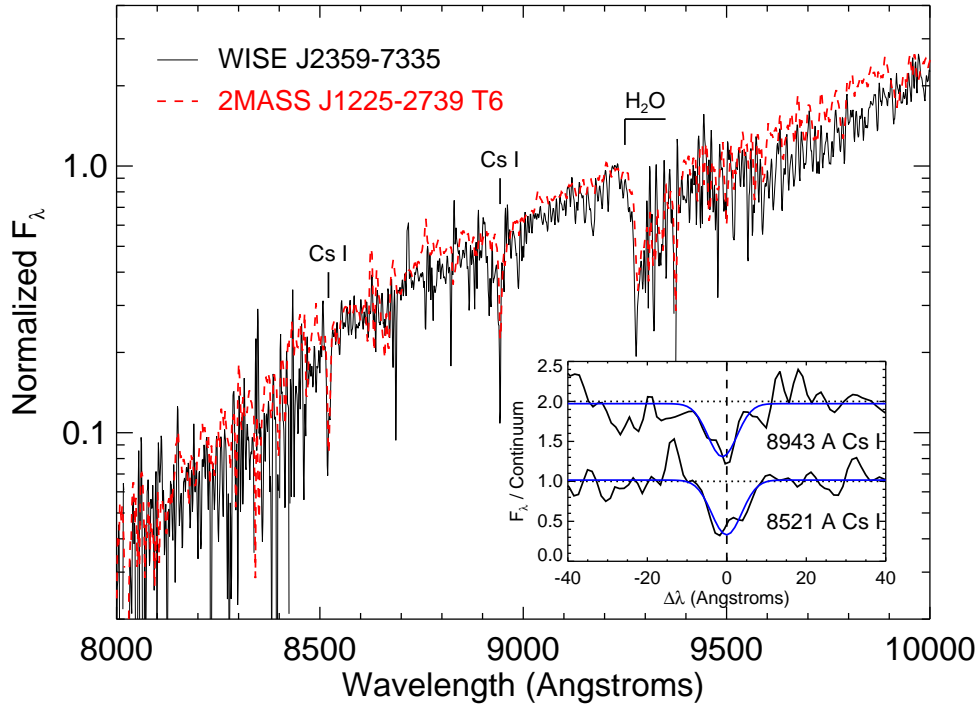}
\caption{LDSS-3 optical spectrum of WISE~J2359$-$7335 (black line), compared to the T6 dwarf 2MASS~J1225$-$2739 (red line; data from \citealt{2003ApJ...594..510B}).  Both spectra are normalized at 9250~{\AA}.  Absorption features from Cs~I and {\wat} are indicated.  The inset box shows a close-up of the Cs~I line profiles after dividing through by the local continuum. Blue lines indicate Gaussian fits to these profiles. The 8943~{\AA} line profile in the inset is offset by a constant for clarity.
\label{fig_opt2359}}
\end{figure*}

Figure~\ref{fig_opt2359} displays a portion of the WISE~J2359$-$7335 
LDSS-3 spectra spanning 8000--10000~{\AA}, compared to equivalent data\footnote{These data 
were obtained with the Low-Resolution Imaging Spectrometer (LRIS; \citealt{1995PASP..107..375O}).} for the T6 dwarf 2MASS~J12255432$-$2739466 (hereafter 2MASS~J1225$-$2739; \citealt{1999ApJ...522L..65B, 2003ApJ...594..510B}).  Both spectra are
logarithmically scaled to highlight absorption features within their steep red optical slopes.
Absorption from Cs~I (8521 and 8943~{\AA}) and {\wat} (9250~{\AA} bandhead)
are visible.  The pseudoequivalent widths of the Cs~I lines were measured to
be 4.9$\pm$1.7~{\AA} and 8.9$\pm$2.5~{\AA}, respectively,
the latter consistent with measurements for mid-type T dwarfs \citep{2003ApJ...594..510B}.

\subsubsection{AAT/IRIS2}

Near-infrared spectroscopy of WISE~J2359$-$7335 was obtained on 2006 May 15 and 2006 June 11 (UT) with the Infrared Imager and Spectrograph (IRIS2; \citealt{2004SPIE.5492..998T}) mounted on the 3.9m Anglo-Australian Telescope (AAT).  Conditions on both nights were clear but humid with poor seeing (1$\farcs$5--2$\arcsec$). 
Spectra in the $J$-band (1.47--1.81~$\micron$) were obtained in May using the 1$\arcsec$-wide slit,  Sapphire-240 transmission grating and J$_l$ filter, at an airmass of 1.49.  Spectra in the $H$-band (1.47--1.81~$\micron$) were obtained in both May and June using the 1$\arcsec$-wide slit,  SAPPHIRE-316 transmission grating and H$_s$ filter, at airmasses of 1.60 and 1.52, respectively.  Average resolution of these spectral modes is {\ldl} = 2100.  Individual exposures of 150~s were obtained in ABBA dither patterns nodding along the slit, for a total exposures of 600~s at $J$ and 1200~s at $H$.  The G0~V star HIP~118079 was observed on both nights immediately after the WISE target for telluric absorption and flux calibration.  Exposures of Quartz halogen and Xe lamps reflected off of the AAT flat-field screen were also obtained at the beginning of each night for pixel response correction and high-order dispersion  
calibration of the wavelength scale. The latter calibration was then updated using the telluric OH emission in the science frames.  Data were reduced following the procedures described in \citet{2005AJ....130.2326T}.  

\begin{figure*}
\centering
\epsscale{1.1}
\plottwo{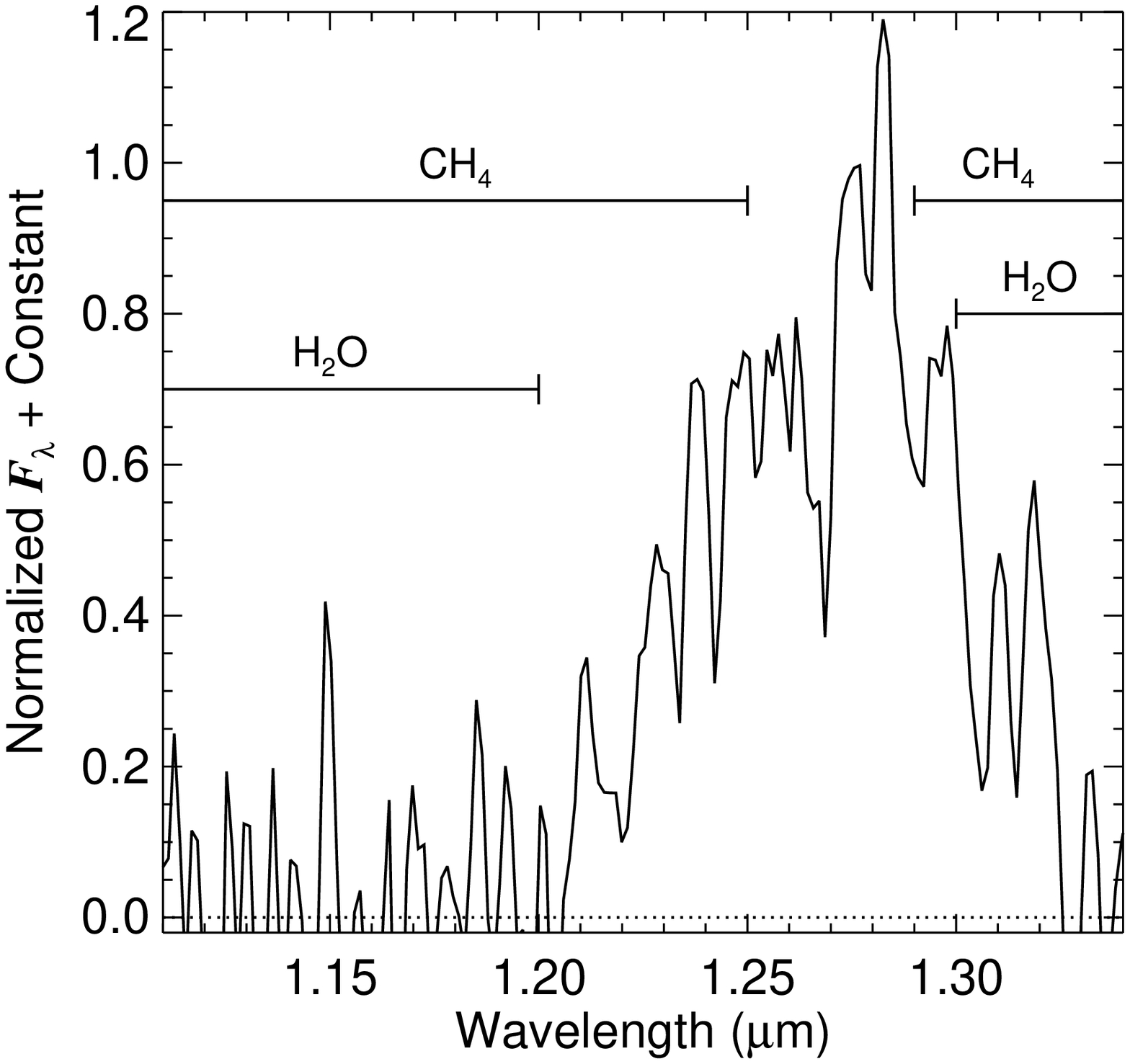}{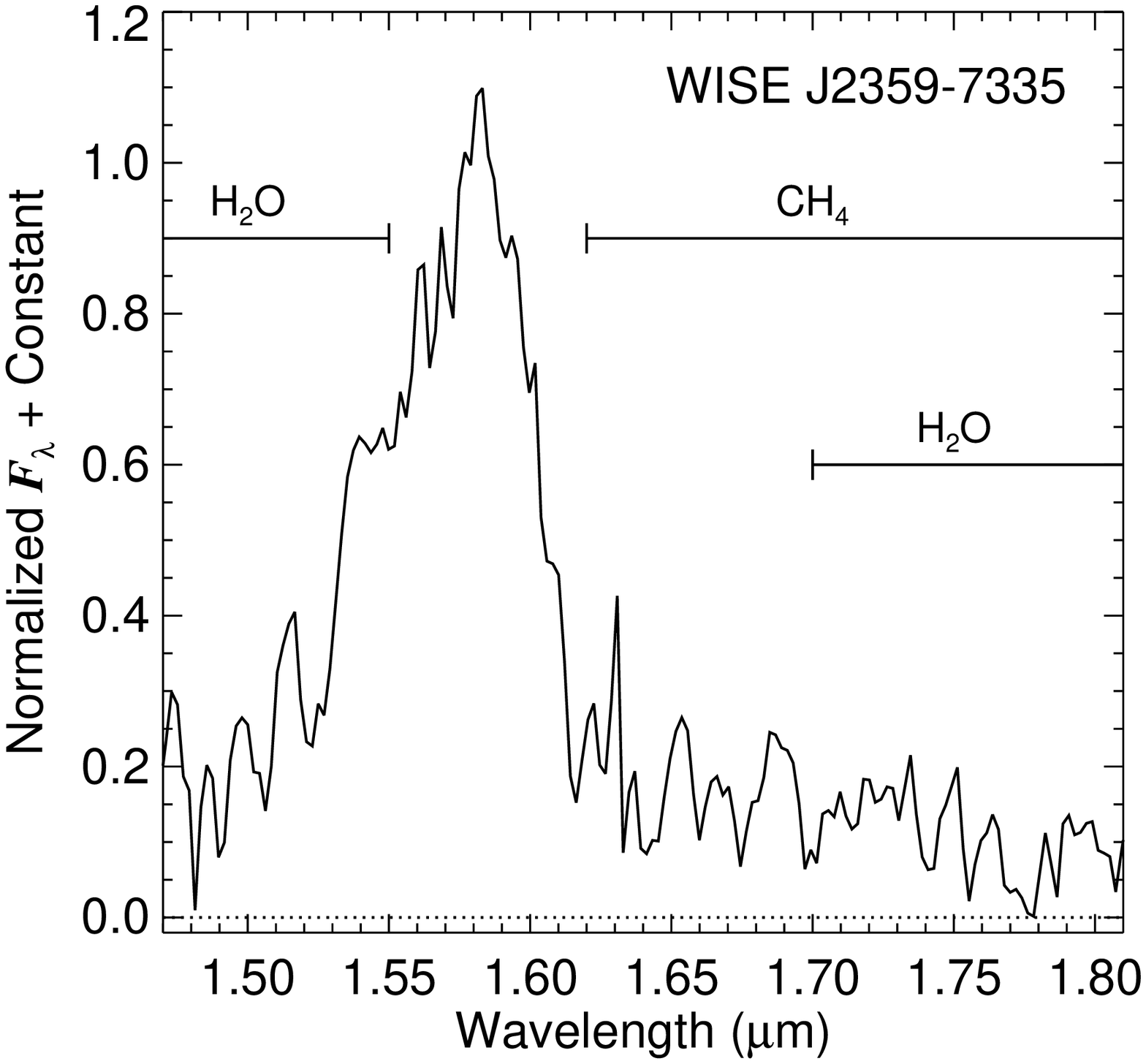}
\caption{IRIS2 $J$- (left) and $H$-band (right) spectra of WISE~J2359$-$7335, smoothed to a resolution of {\ldl} = 500 using a gaussian kernel.  Data are normalized at their respective flux peaks.  Primary absorption bands from {\wat} and {\meth} are indicated.
\label{fig_iris2}}
\end{figure*}

Figure~\ref{fig_iris2} displays the resulting spectra, smoothed to {\ldl} = 500.  The data have relatively low signal-to-noise ($\sim$5 at the 1.27~$\micron$ peak; $\sim$10 at the 1.58~$\micron$ peak), but are nevertheless sufficient to resolve the strong near-infrared {\wat} and {\meth} bands characteristic of mid- to late-type T dwarf spectra.

\subsubsection{Magellan/FIRE}

Near-infrared spectroscopy of the WISE candidates and the T8.5 dwarf companion brown dwarf 
Wolf~940B \citep{2009MNRAS.395.1237B}
was obtained on three separate runs during 2010 April 7,  2010 September 19-20,
and 2010 December 24 (UT).
All targets were observed with FIRE in its low-resolution, prism-dispersed mode,
which delivers 0.85--2.45~$\micron$ continuous spectroscopy in a single order.  
Each source was initially acquired using FIRE's $J$-band acquisition camera, then
placed into a 0$\farcs$6 slit aligned to the parallactic angle (rotator angle 89$\fdg$5).
This prism/slit combination provides a variable resolution of {\ldl} = 250-350 across
the near-infrared band.
A series of AB or ABBA dither exposure sequences were obtained with integrations ranging from
60~s to 120~s per exposure (plus 10.6~s read time), the latter being the maximum permitted 
to avoid saturating OH telluric lines in the $H$-band.
The spectrograph detector was read out using the 4-amplifier mode at ``high gain'' 
(1.2~counts/e$^-$) with either Fowler-8 sampling (April and September) or Sample Up The Ramp 
(December) modes.
Each science target observation was accompanied by an observation of a 
nearby A0~V calibrator star (typically with $V$ = 10--12) at a similar airmass.
Given FIRE's high sensitivity, these calibrators were occasionally observed out of focus and/or offset from the slit to avoid saturation in FIRE's 
minimum readout time (11.6~s).
We also obtained exposures of a variable voltage quartz flat field lamp (set at 1.2~V and 2.2~V)
and arc lamps (NeAr) reflected off of Baade's secondary flat-field screen for pixel response and wavelength calibration.
Data were reduced using a combination of IRAF (NOAO {\it onedspec} package) and IDL\footnote{Interactive Data Language.} routines (SpeXtool {\it xcombspec} and {\it xtellcor{\_}general}; \citealt{2003PASP..115..389V, 2004PASP..116..362C}), as described in detail in \citet{2010ApJ...725.1405B}.  

\begin{figure*}
\centering
\epsscale{0.85}
\plotone{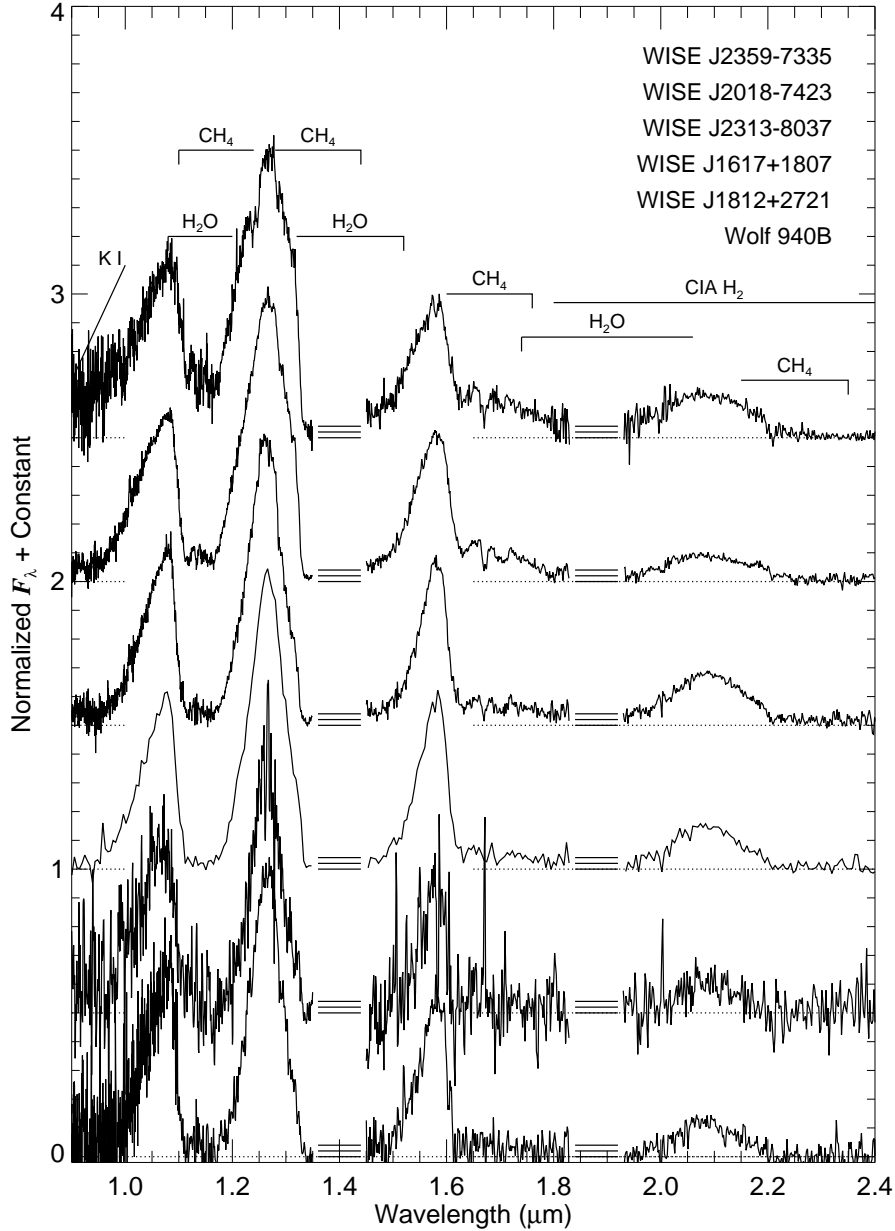}
\caption{FIRE prism spectra of WISE T dwarf discoveries (in order of spectral morphology) and the T8.5 comparison object Wolf~940B.  
Spectra are normalized at their 1.3~$\micron$ spectral peaks and vertically offset by a constant (dotted lines).
Regions of strong telluric absorption (1.35--1.45~$\micron$ and 1.82--1.95~$\micron$)
have been excised from the plot.
Major spectral features characteristic of T dwarf spectra are labeled. 
\label{fig_spec_wise}}
\end{figure*}

Figure~\ref{fig_spec_wise} displays the reduced FIRE spectra.  
Signal-to-noise at the 1.27~$\micron$ peak varies from $\sim$20 for WISE~J1812+2721 to $\sim$100 for WISE~J1617+1807.
All of the WISE spectra exhibit strong {\wat} and {\meth} bands, absorption from the pressure-broadened red wing of the 0.77~$\micron$ K~I doublet, 
and blue near-infrared spectral energy distributions, characteristic signatures of late-type T dwarfs.
The 1.1~$\micron$ and 1.6~$\micron$ bands are particularly deep
in the spectra of WISE~J1617+1807, WISE~J1812+2721 and WISE~J2313$-$8037, comparable to the 
bands seen in the spectrum of Wolf 940B.
These features are weakest in the spectrum of WISE~J2359$-$7335.
WISE~J2018$-$7423 exhibits an unusually flat 2.1~$\micron$ $K$-band flux peak compared to the other sources, 
a region dominated by collision-induced {\hh} absorption (\citealt{1969ApJ...156..989L, 1994ApJ...424..333S}; see Section~6.3).  
Note that the lower signal-to-noise spectrum of WISE~J1812+2721 is due to its faintness ($J$ = 18.19$\pm$0.02) and poor observing conditions on 2010 September 19 (airmass $\sim$ 2; seeing $\gtrsim$ 1$\farcs$5).  Nevertheless, its overall spectral shape is clearly indicative of a very late-type T dwarf.

\section{Characterizing the T Dwarfs}

\subsection{Spectral Classification}

\begin{figure*}
\centering
\epsscale{1.0}
\includegraphics[width=0.4\textwidth]{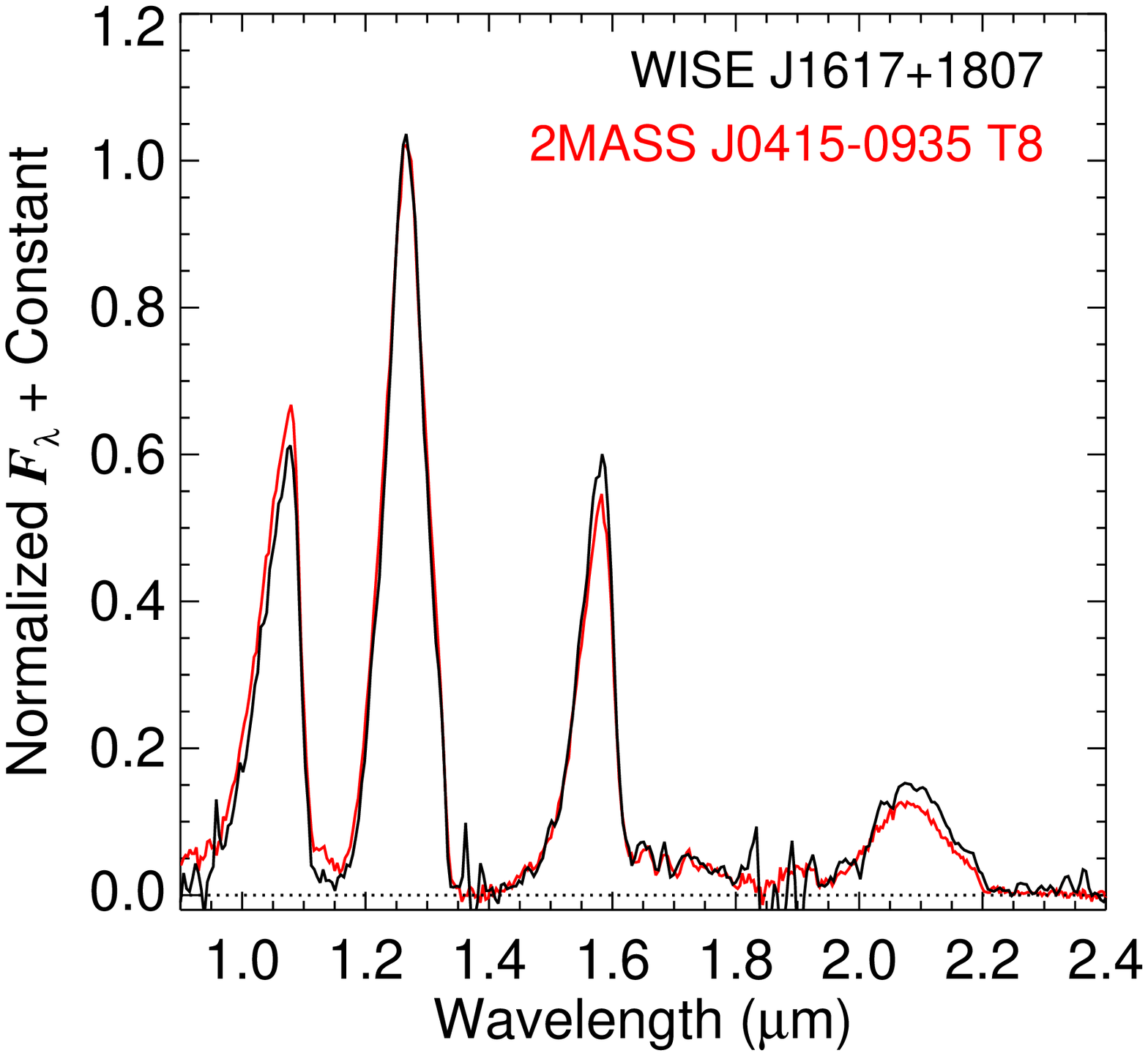}
\includegraphics[width=0.4\textwidth]{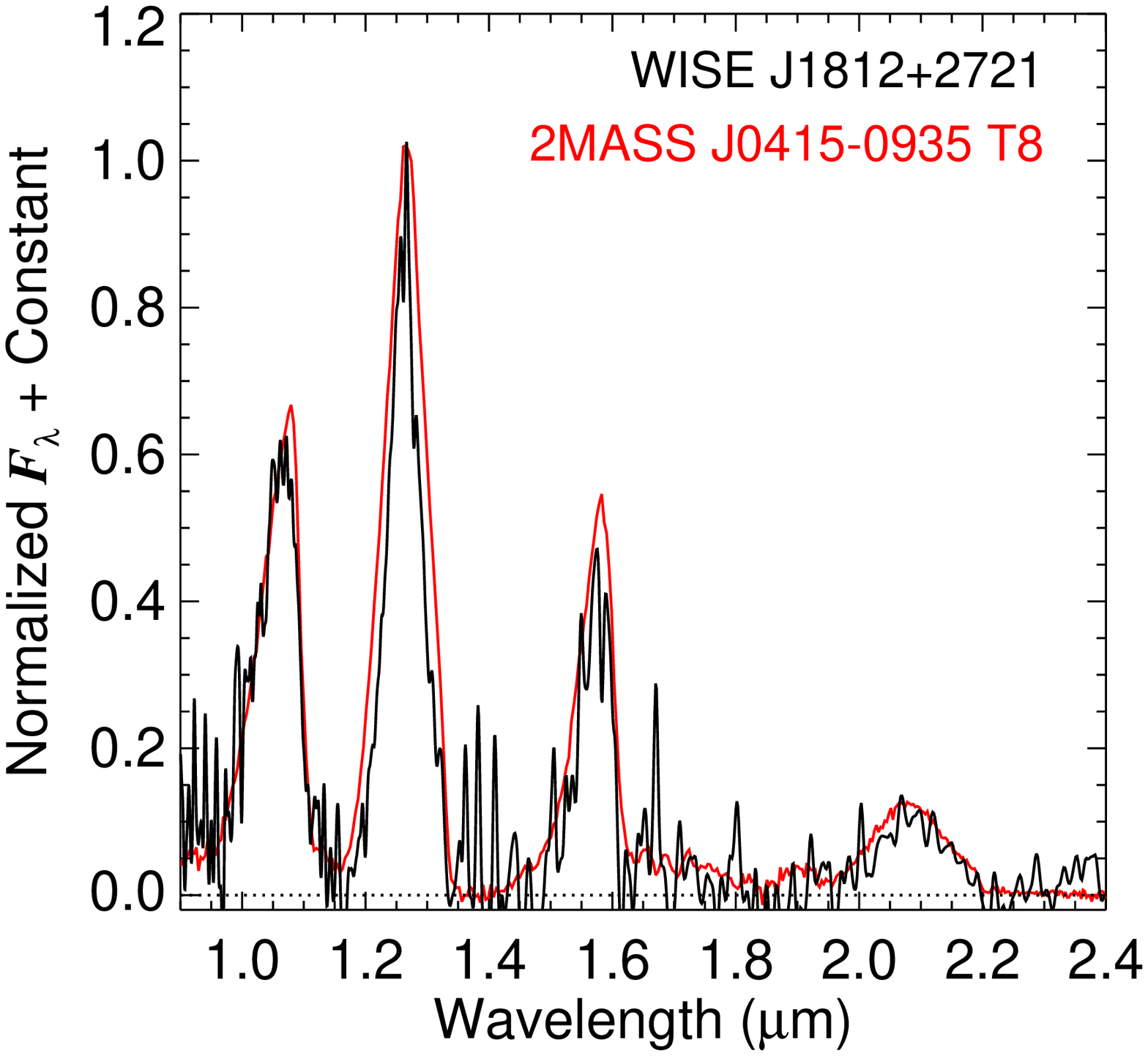}
\includegraphics[width=0.4\textwidth]{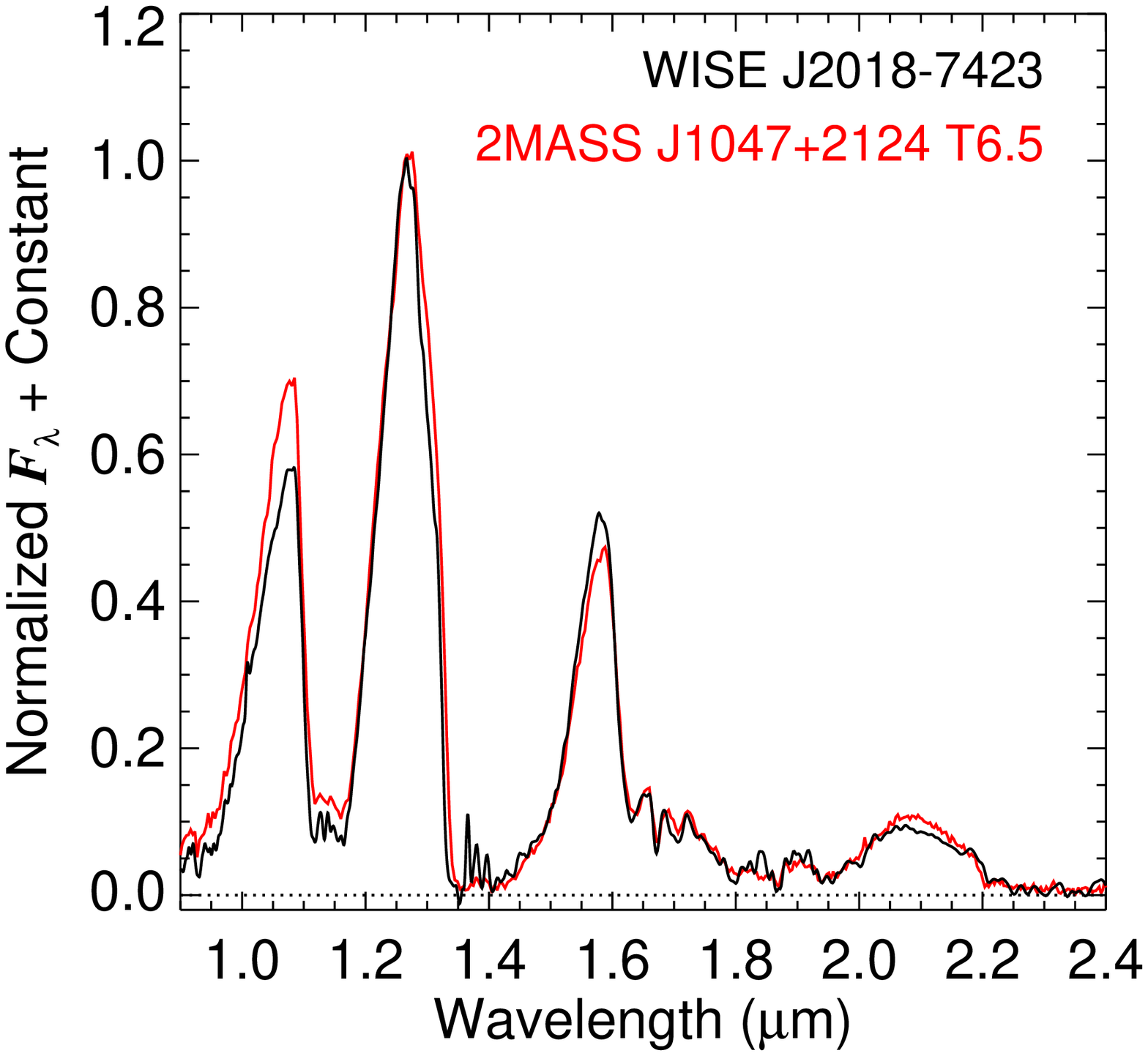}
\includegraphics[width=0.4\textwidth]{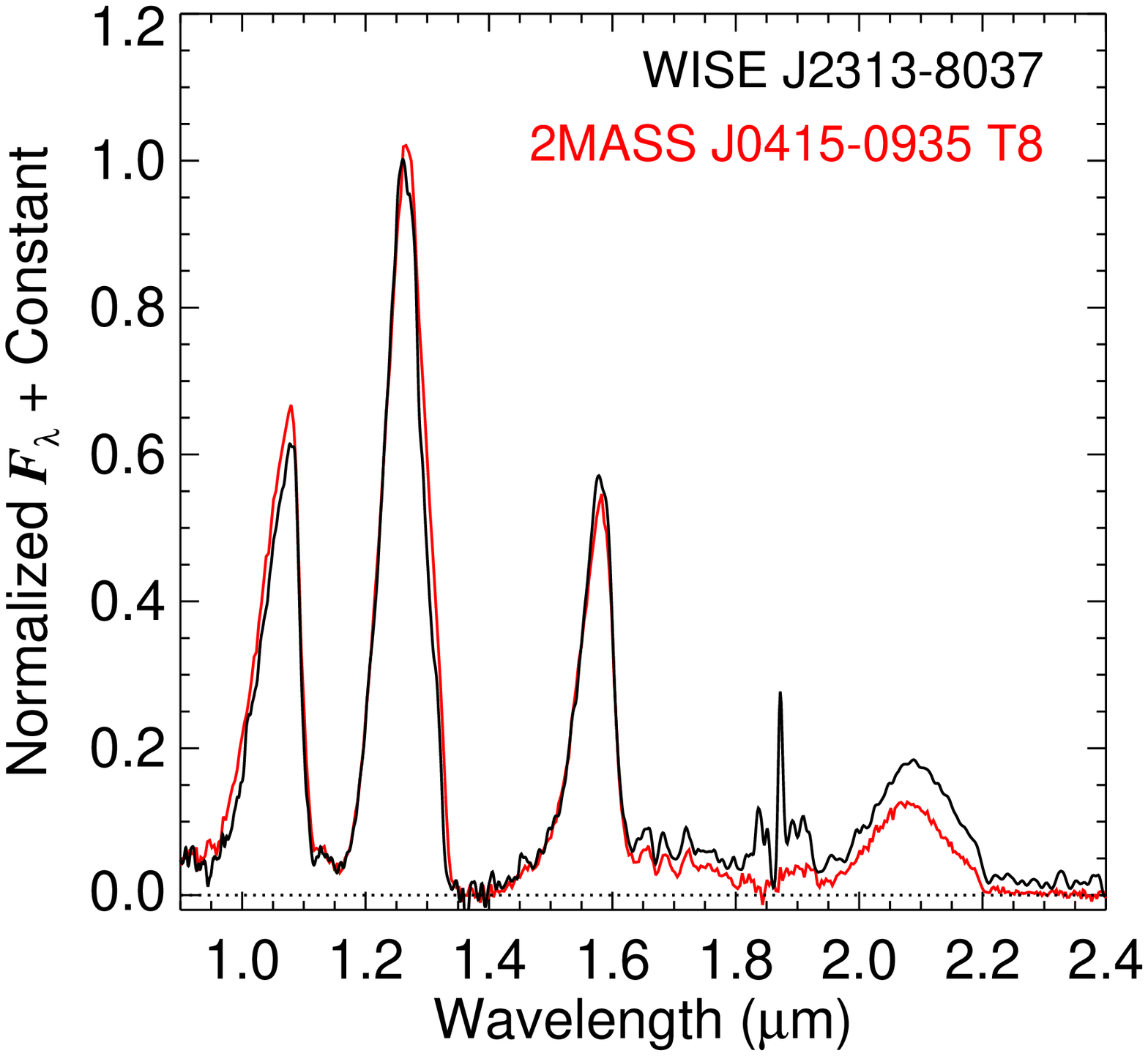}
\includegraphics[width=0.4\textwidth]{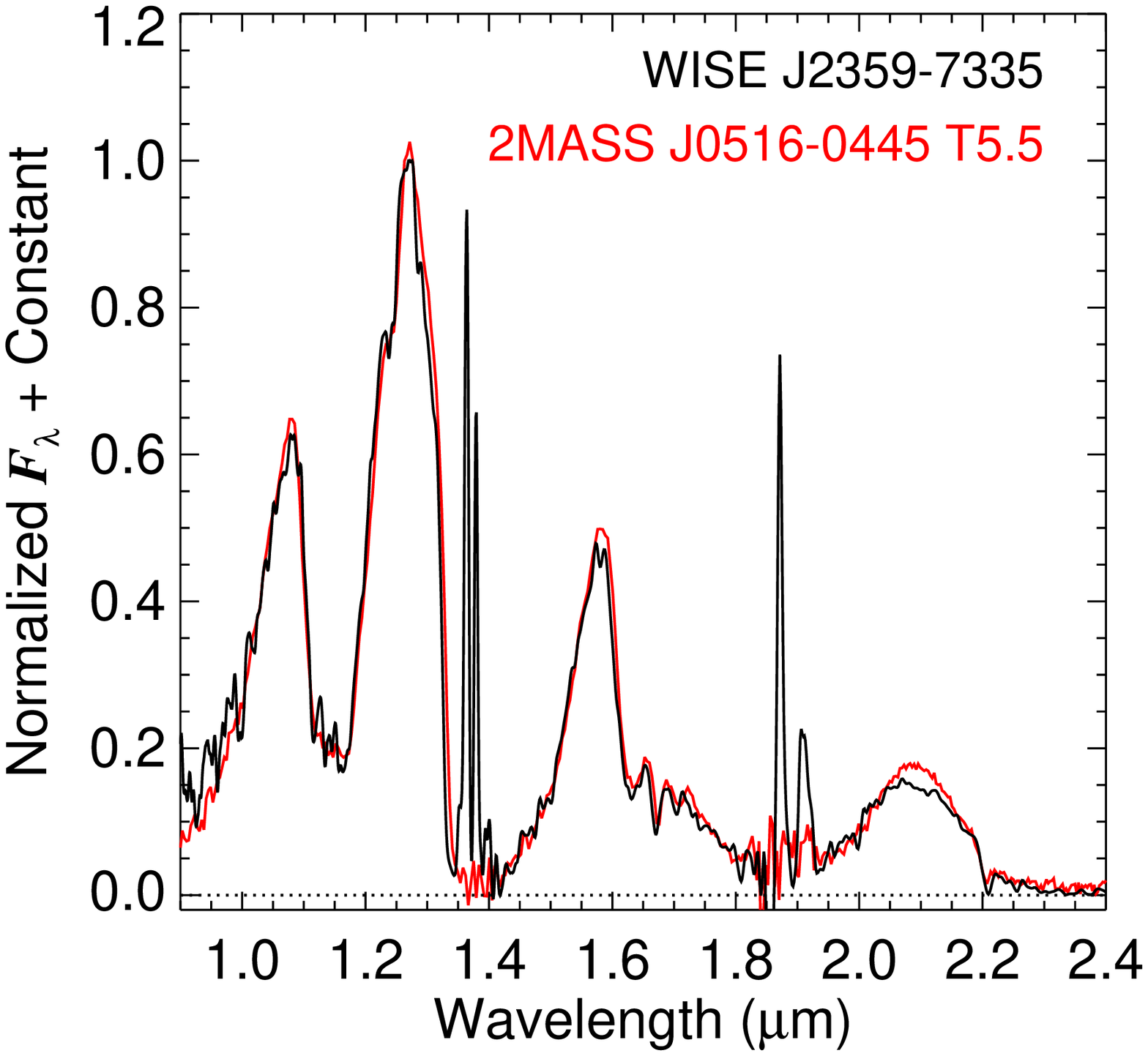}
\caption{Individual FIRE spectra of the WISE T dwarfs (black lines) compared to their best-fitting SpeX spectral templates (red lines): 2MASS~J04151954$-$0935066 (T8; \citealt{2002ApJ...564..421B, 2004AJ....127.2856B}), 
2MASS~J10475385+2124234 (T6.5; \citealt{1999ApJ...522L..65B, 2008ApJ...681..579B})
and 2MASS~J05160945$-$0445499 (T5.5; \citealt{2003AJ....126.2487B, 2008ApJ...681..579B}).
All spectra are normalized at the 1.3~$\micron$ flux peaks, and the FIRE data have been smoothed to match the 
resolution of the SpeX data ({\ldl} $\approx$ 120) using a gaussian kernel. 
\label{fig_class}}
\end{figure*}

The T dwarfs were classified using their FIRE spectra following two methods.  First, we compared the data to a suite of spectral templates drawn from the SpeX Prism Spectral Libraries\footnote{See \url{http://www.browndwarfs.org/spexprism}.}, including the T dwarf standards defined in \citet{2006ApJ...637.1067B}.  The SpeX prism data \citep{2003PASP..115..362R} have lower resolution than the FIRE data, {\ldl} = 90--120; we therefore smoothed the latter to this resolution using a gaussian kernel.  We quantified the agreement between normalized WISE and template spectra using the $\chi^2$ statistic, sampling over the wavelength regions 1.0--1.35~$\micron$, 1.45--1.8~$\micron$ and 2.0--2.4~$\micron$ to avoid strong telluric absorption.  Figure~\ref{fig_class} displays the best-matching templates for each of the WISE targets.  Note that the spectra of WISE~J1812+2721 and (to a lesser extent) WISE~J2313$-$8037 appear later than that of the T8 spectral standard 2MASS~J04151954$-$0935066 (hereafter 2MASS~J0415$-$0935; \citealt{2002ApJ...564..421B}) based on their narrower $J$-band flux peaks.

\begin{deluxetable*}{lcccccc}
\tabletypesize{\scriptsize}
\tablecaption{Spectral Indices for Observed T Dwarfs. \label{tab_indices}}
\tablewidth{0pt}
\tablehead{
\colhead{Index} &
\colhead{WISE J1617+1807} &
\colhead{WISE J1812+2721} &
\colhead{WISE J2018-7423} &
\colhead{WISE J2313-8037} &
\colhead{WISE J2359-7335} &
\colhead{Wolf 940B} \\
}
\startdata
{\wat}-J & 0.020$\pm$0.004 ($\geq$T8) & 0.04$\pm$0.05 (T7/$\geq$T8) & 0.086$\pm$0.006 (T7) & 0.044$\pm$0.008 ($\geq$T8) & 0.206$\pm$0.008 (T5) & 0.026$\pm$0.012 ($\geq$T8) \\
{\meth}-J & 0.170$\pm$0.003 ($\geq$T8) & 0.12$\pm$0.03 ($\geq$T8) & 0.198$\pm$0.004 ($\geq$T8) & 0.116$\pm$0.004 ($\geq$T8) & 0.261$\pm$0.004 (T7) & 0.100$\pm$0.013 ($\geq$T8) \\
{\wat}-H & 0.159$\pm$0.007 (T8) & 0.14$\pm$0.13 (T6/$\geq$T9) & 0.244$\pm$0.011 (T7) & 0.174$\pm$0.012 (T8) & 0.343$\pm$0.011 (T5) & 0.13$\pm$0.04 (T8/$\geq$T9) \\
{\meth}-H & 0.108$\pm$0.006 ($\geq$T8) & 0.29$\pm$0.11 (T5/T7) & 0.231$\pm$0.010 (T7) & 0.142$\pm$0.011 ($\geq$T8) & 0.315$\pm$0.008 (T6) & 0.08$\pm$0.04 ($\geq$T8) \\ 
{\meth}-K & 0.033$\pm$0.016 ($\geq$T7) & -0.03$\pm$0.23 (N/A) & 0.17$\pm$0.04 (T5/$\geq$T7) & 0.14$\pm$0.03 (T6/$\geq$T7) & 0.145$\pm$0.007 (T6) & 0.00$\pm$0.12 (N/A) \\
$W_J$ & 0.275$\pm$0.003 (T8/$\geq$T9) & 0.22$\pm$0.04 ($\geq$T9) & 0.420$\pm$0.005 ($\leq$T6) & 0.324$\pm$0.005 (T8) & 0.549$\pm$0.005 ($\leq$T6) & 0.251$\pm$0.014 ($\geq$T9) \\
K/J & 0.156$\pm$0.002 & 0.14$\pm$0.03 &  0.097$\pm$0.003 & 0.196$\pm$0.004 & 0.158$\pm$0.002 & 0.135$\pm$0.012 \\
$J-K$\tablenotemark{a} & -0.21$\pm$0.05 & -0.5$\pm$0.8 & -0.54$\pm$0.10 &  0.12$\pm$0.07 &  -0.48$\pm$0.02 &  -0.7$\pm$0.4 \\
Template SpT & T8 & $\geq$T8 & T6.5 &  $\geq$T8 &  T5.5 & $\geq$T8 \\
Adopted SpT & T8 & T8.5: & T7 & T8 & T5.5 & T8.5 \\
\enddata
\tablecomments{Index spectral types based on the index ranges defined in \citet{2010MNRAS.406.1885B},
which incorporates the definitions set out by \citet{2006ApJ...637.1067B} for T0-T8 dwarfs and \citet[for the $W_J$ index]{2008MNRAS.391..320B} for T9 dwarfs.  The final type is an average of the index types and the template classification, accounting for upper/lower limits.}
\tablenotetext{a}{Spectrophotometric colors computed from the spectral data following \citet{2005ApJ...623.1115C}.} 
\end{deluxetable*}

We also computed the near-infrared classification indices {\wat}-J, {\meth}-J, {\wat}-H, {\meth}-H, {\meth}-K and $W_J$ from the FIRE data using the definitions given in \citet{2006ApJ...637.1067B} and \citet{2007MNRAS.381.1400W}, and the spectral type/index ranges defined
in \citet{2010MNRAS.406.1885B} which extend to type T9.
For completeness, we also measured the K/J index defined in \citet{2006ApJ...637.1067B}
and the spectrophotometric $J-K$ color on the MKO\footnote{Mauna Kea Observatory filter system; see \citet{2002PASP..114..180T} and \citet{2002PASP..114..169S}.} system following \citet{2005ApJ...623.1115C}.
We accounted for uncertainty in these measures through Monte Carlo simulation, sampling 1000 realizations of each spectrum varied pixel-by-pixel by random offsets drawn from a normal distribution scaled to the noise spectrum.  The final index values, listed in Table~\ref{tab_indices}, reflect the means and standard deviations of these measurements.
The associated spectral types for each index, rounded off to the nearest half subtype, are also listed in Table~\ref{tab_indices}.
These types are generally in agreement with each other and with the template-comparison classification,
although the noisier spectrum of WISE~J1812+2721 results in greater scatter.  
The final classifications were taken as an average of the index and template classifications, accounting for limits in the index types.  Classifications range from T5.5 for WISE~J2359$-$7335 to T8.5: for WISE~J1812+2721, where the colon indicates an uncertain classification due to noise.  WISE~J1617+1807, WISE~J1812+2721 and WISE~J2313$-$8037 are all classified as T8 and later.

\subsection{Estimated Distances and Kinematics}

To estimate the distances of these T dwarfs, we first derived a linear absolute $W2$ magnitude/spectral type relation for T6--T8 dwarfs based on WISE photometry \citep{2011ApJ...726...30M} and parallax measurements \citep{1997A&A...323L..49P, 2003AJ....126..975T, 2004AJ....127.2948V} for the T6 dwarf SDSSp~J162414.37+002915.6 \citep{1999ApJ...522L..61S}, 
the T7.5 dwarf Gliese~570D \citep{2000ApJ...531L..57B} 
and the T8 dwarf 2MASS~J0415$-$0935.
The inferred relation is
\begin{equation}
M_{W2} = 11.33 + 0.268 \times SpT
\end{equation}
where SpT(T6) = 6, SpT(T8) = 8, etc. The scatter in the fit is formally 0.03~mag; however, due to the small number of calibrators used we assume a systematic uncertainty of 0.1~mag.  Distances, taking into account uncertainties in the photometry, spectral classification (0.5-1.0 subtypes) and absolute magnitude relation are listed in Table~\ref{tab_characteristics}.  All of the WISE T dwarfs in this sample are roughly 12--13~pc from the Sun (modulo 1.5--3~pc uncertainties), assuming they are single. 

For WISE~J2018$-$7423, WISE~J2313$-$8037 and WISE~J2359$-$7335, we combined these distances with proper motion measurements to infer tangential velocities.  We find {\vtan} = 56$\pm$6~{\kms}, 31$\pm$5~{\kms} and 20$\pm$3~{\kms} for these sources, respectively.  The motions of  WISE~J2313$-$8037 and WISE~J2359$-$7335 are consistent with the mean kinematics of nearby field T dwarfs (30$\pm$20~{\kms}; \citealt{2009AJ....137....1F}), while WISE~J2018$-$7423 is a $\gtrsim$1$\sigma$ outlier. This high velocity source is discussed in further detail in Section~6.3.

\section{Spectral Model Fits}

To further characterize these brown dwarfs, we 
compared our FIRE spectra to both cloudy and cloud-free
atmosphere models from  \citet{2008ApJ...689.1327S}.
We restricted our analysis to the near-infrared spectra alone; i.e., we did not include the WISE photometry.  As such, this analysis should be regarded as a 
preliminary reconnaissance of the atmospheric and physical properties of these dwarfs.
A more comprehensive modeling effort will be presented in a forthcoming paper (M.\ Cushing et al.\ 2011, in preparation)

\begin{figure*}
\centering
\epsscale{1.0}
\includegraphics[width=0.9\textwidth]{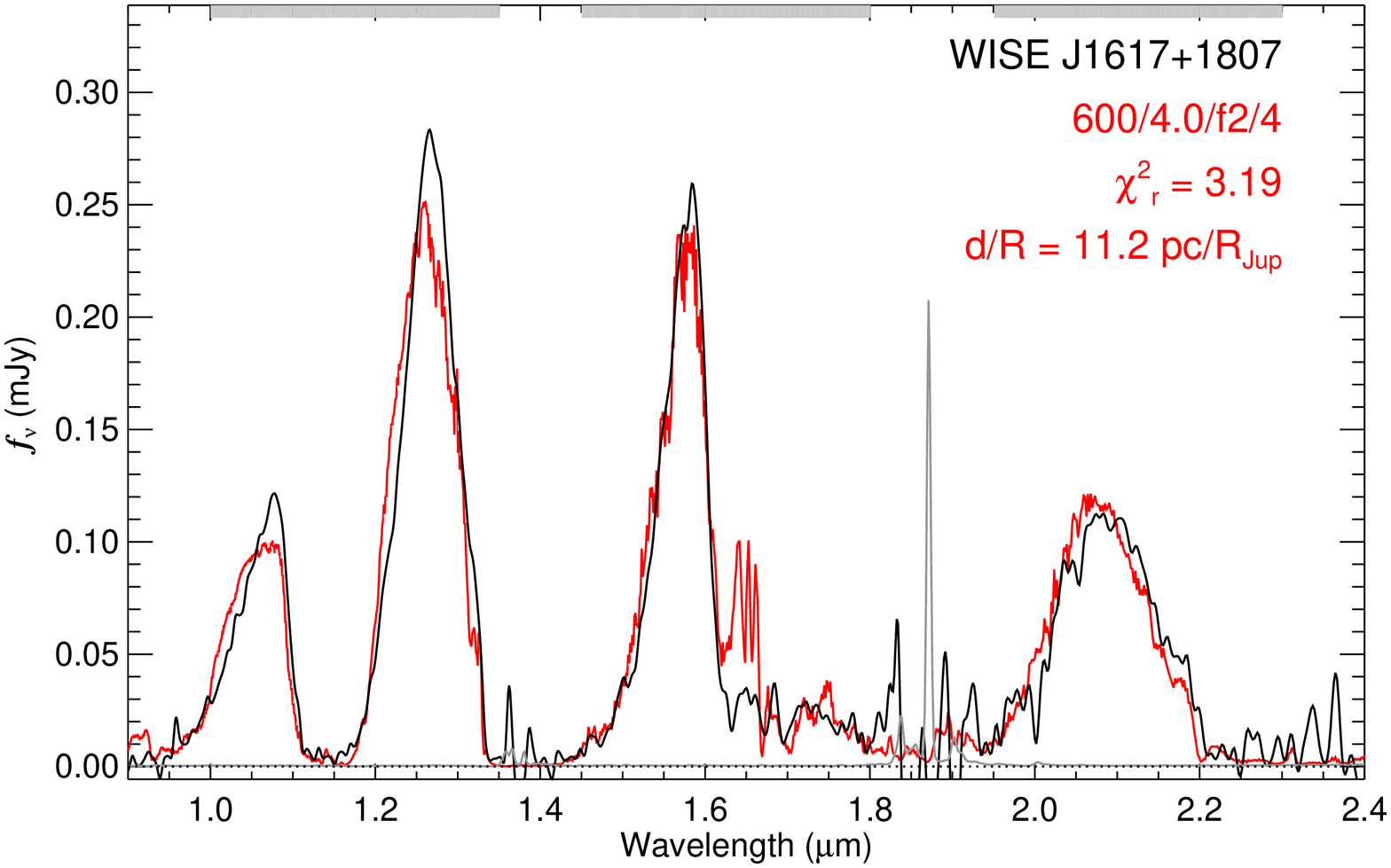}
\includegraphics[width=0.3\textwidth]{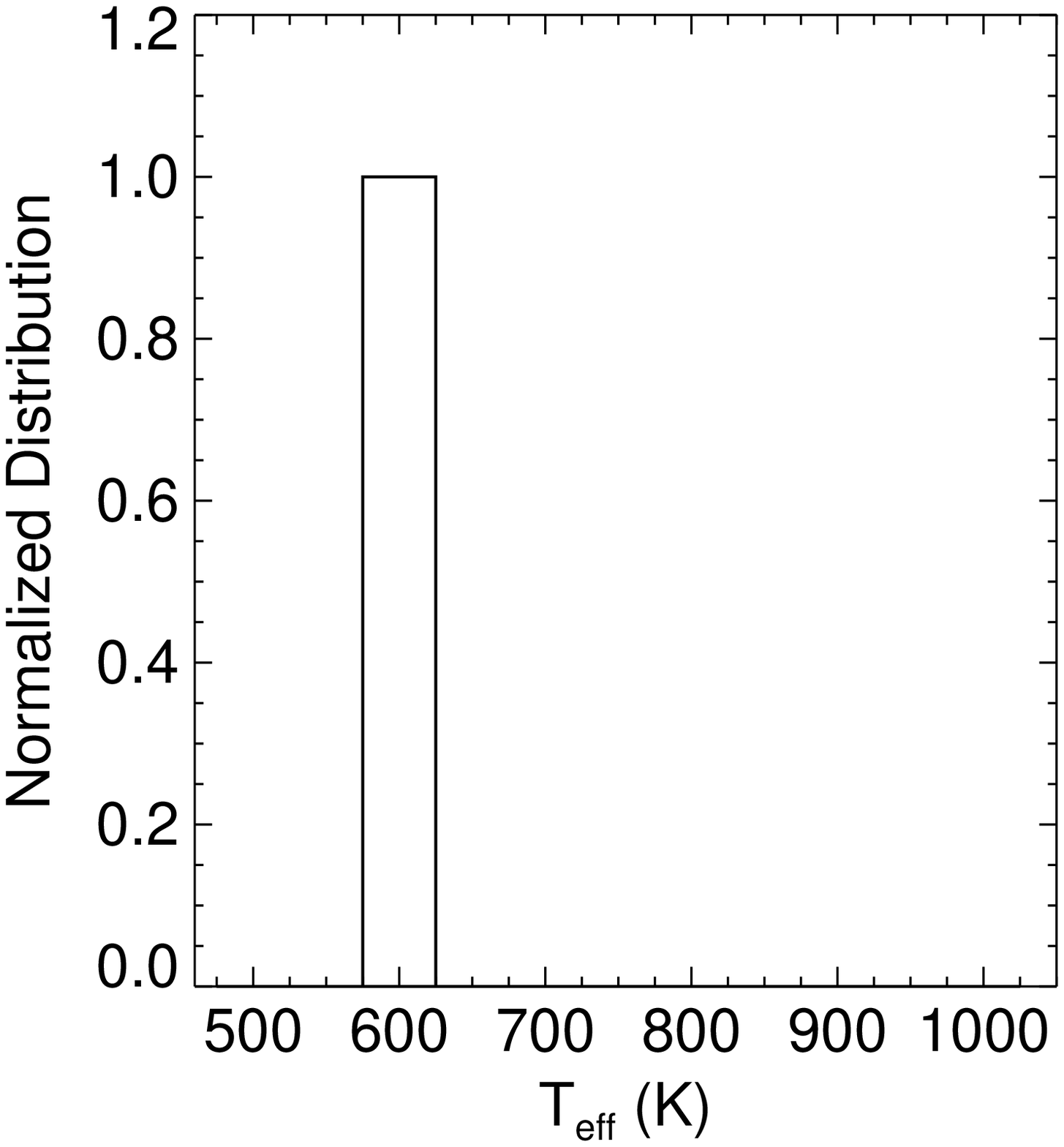}
\includegraphics[width=0.3\textwidth]{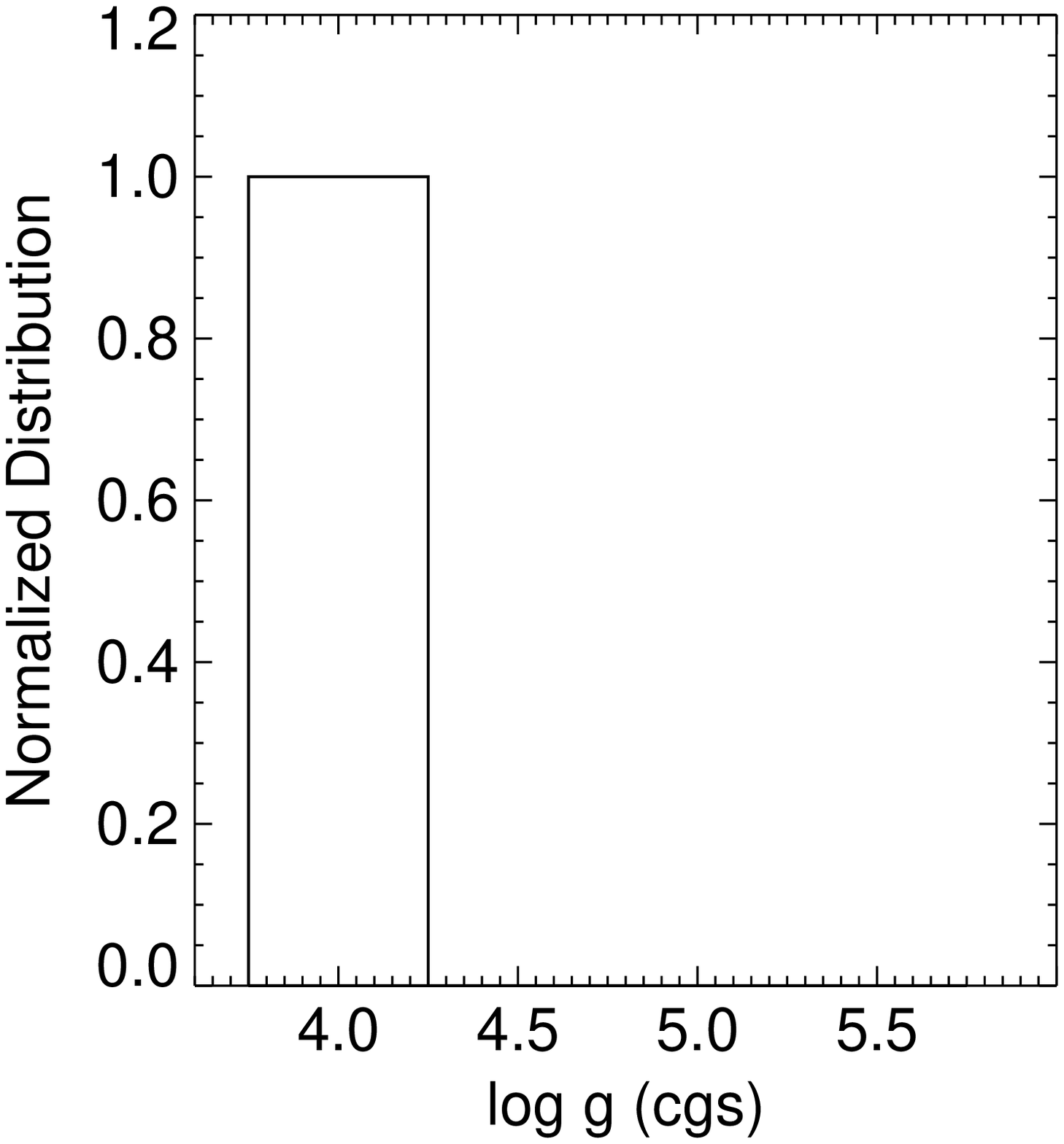}
\includegraphics[width=0.3\textwidth]{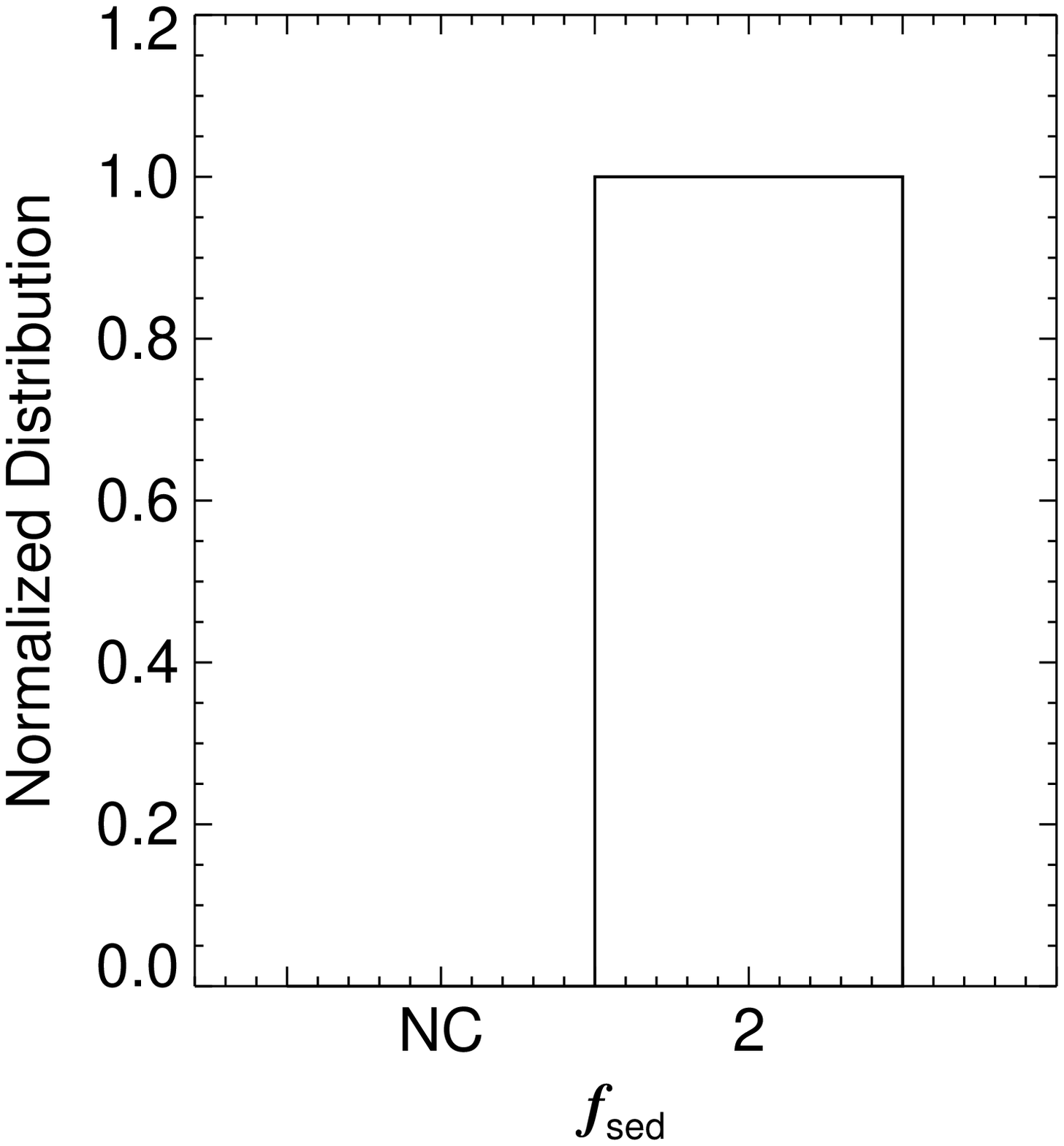}
\caption{(Top panel) Best-fitting spectral model (red line) to FIRE data for WISE~J1617+1807 (black line).
Both spectra are smoothed to the average resolution of the FIRE prism mode ({\ldl} $\approx$ 300). The data are shown in $f_{\nu}$ units 
scaled to the apparent $J$-band magnitude of WISE~J1617+1807, and the model
scaled to minimize $\chi^2$ (the reduced $\chi^2$ is listed).  Model parameters in the form 
{\teff}/{\logg}/{\fsed}/{\logkzz} are listed, with units as given in the text.
We also list the inferred distance-to-radius ratio for this model based on the optimal scaling.
Spectral regions over which the fits were made are indicated by the grey bars at top.
(Bottom panels) From left to right, distributions of {\teff}, {\logg} and {\fsed} based on an F-test PDF factor weighting of each model fit relative to the best-fitting model (see \citealt{2010ApJ...725.1405B}).
\label{fig_fit_1617}}
\end{figure*}

\begin{figure*}
\centering
\epsscale{1.0}
\includegraphics[width=0.9\textwidth]{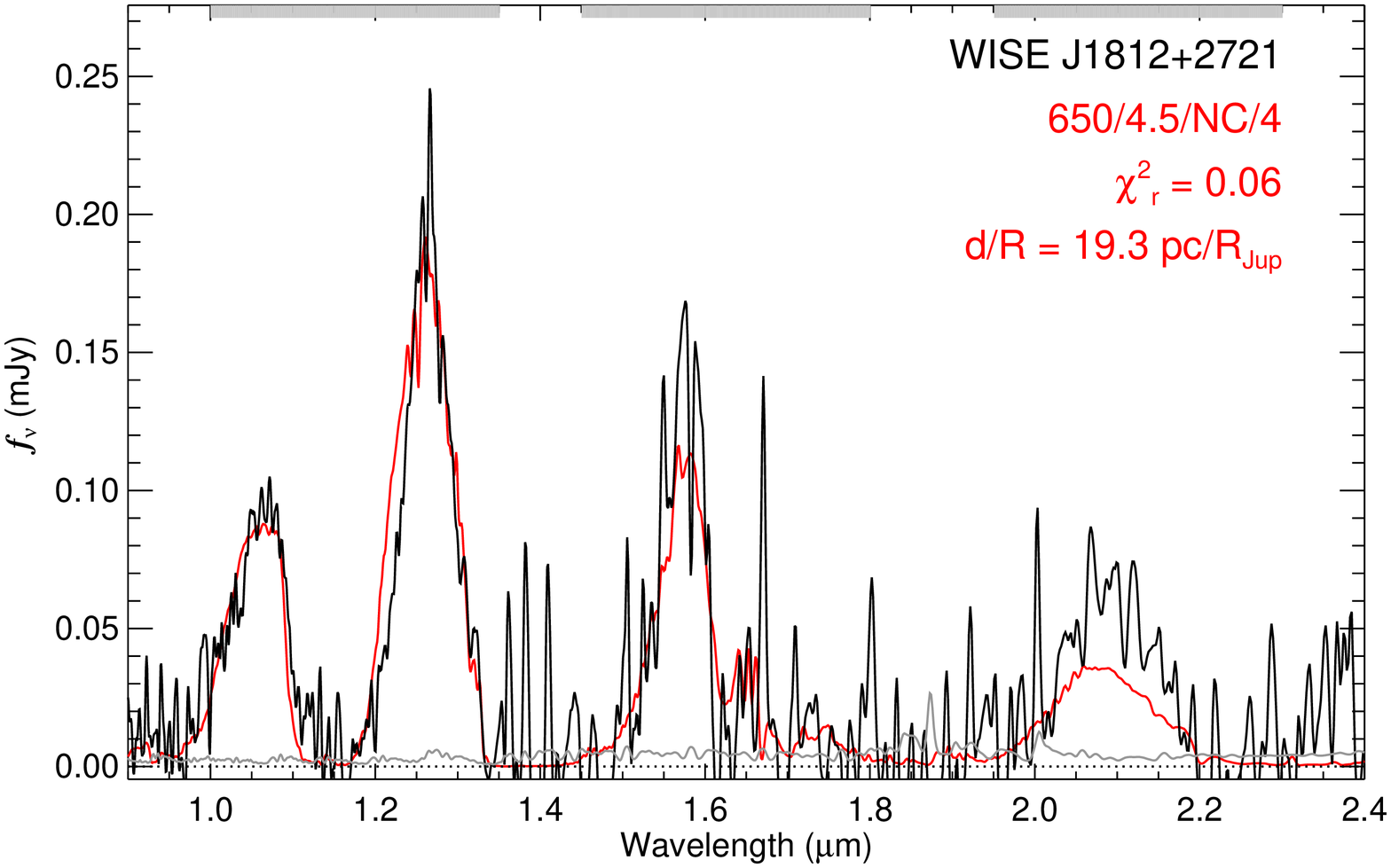}
\includegraphics[width=0.3\textwidth]{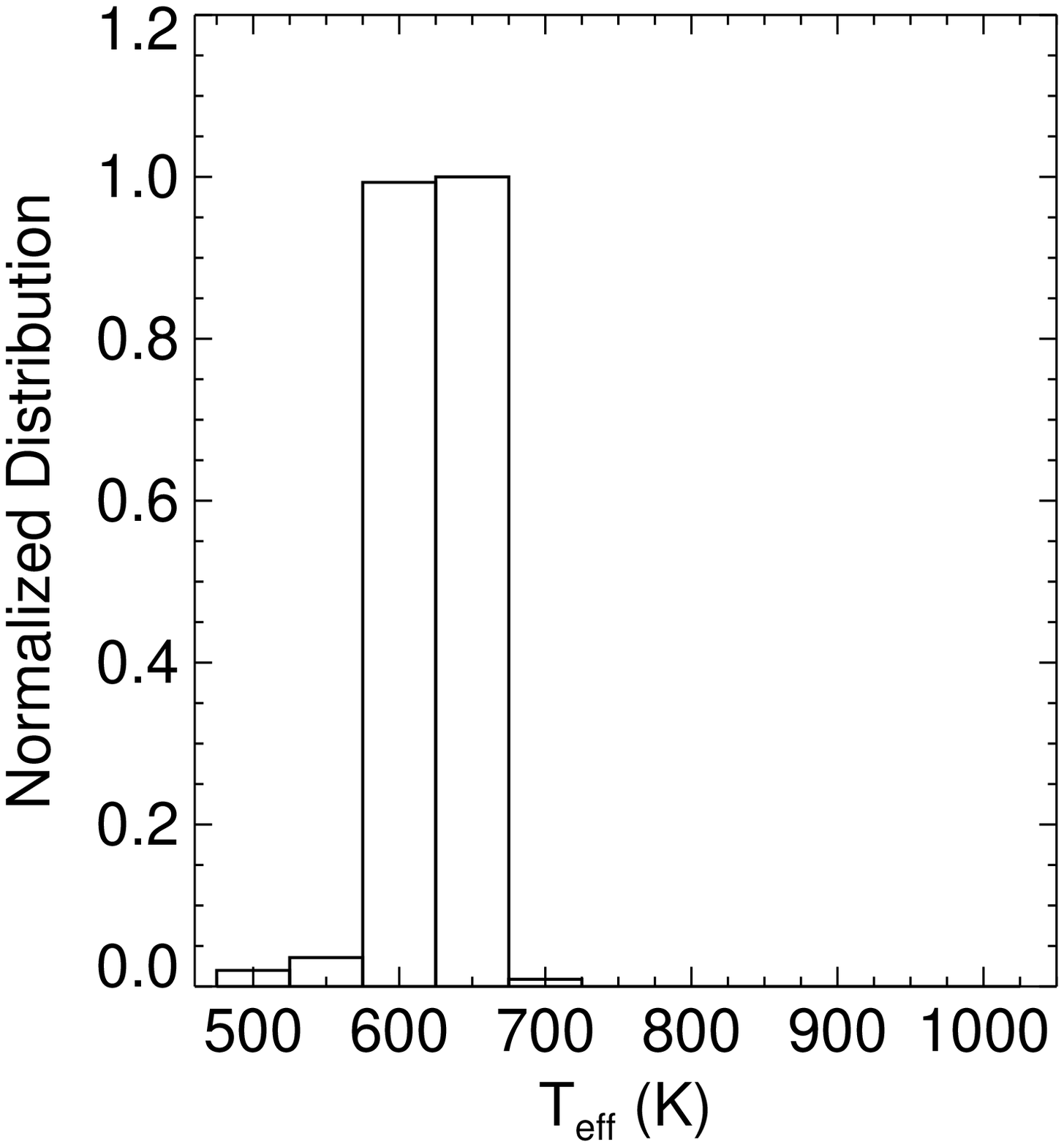}
\includegraphics[width=0.3\textwidth]{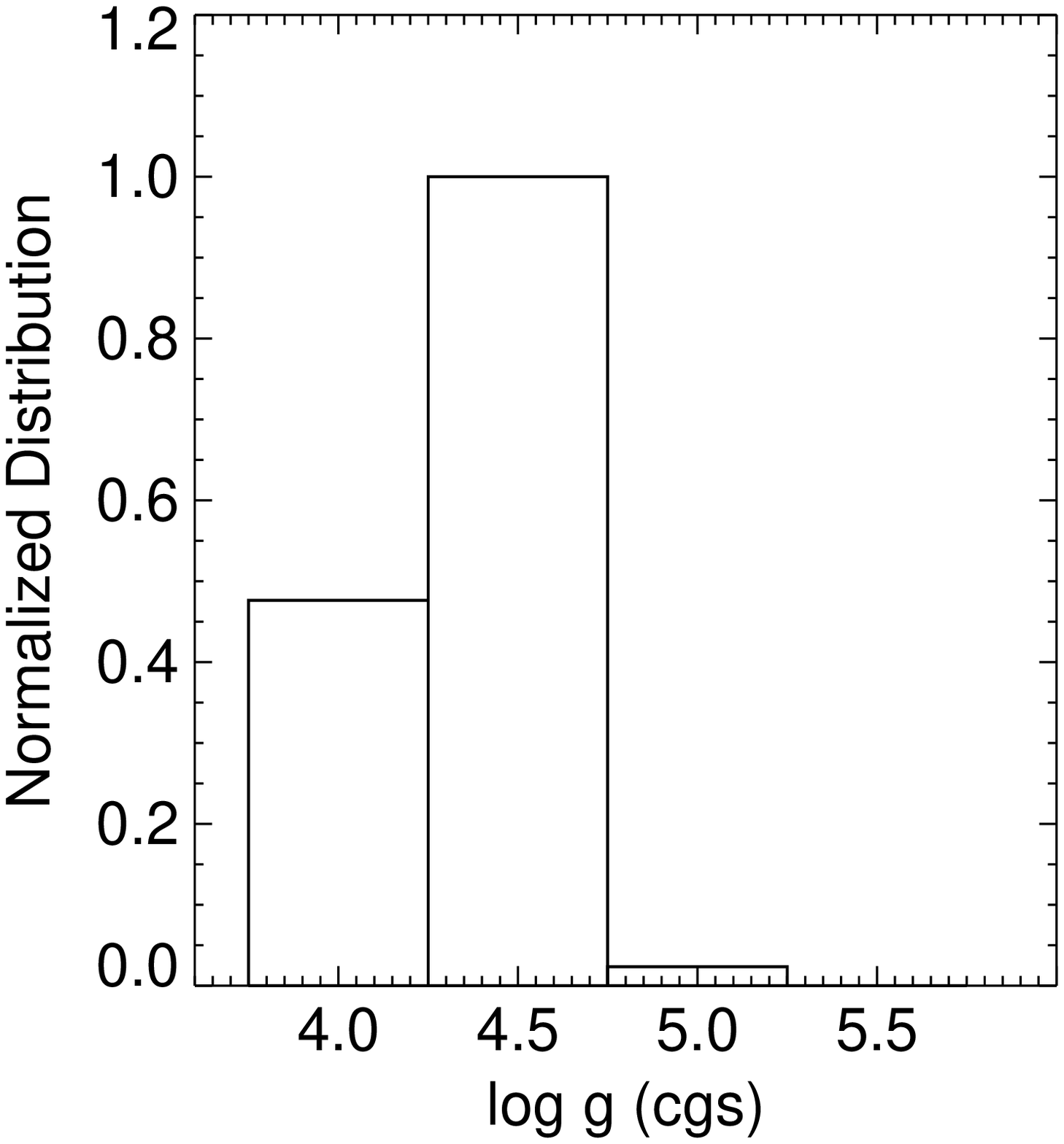}
\includegraphics[width=0.3\textwidth]{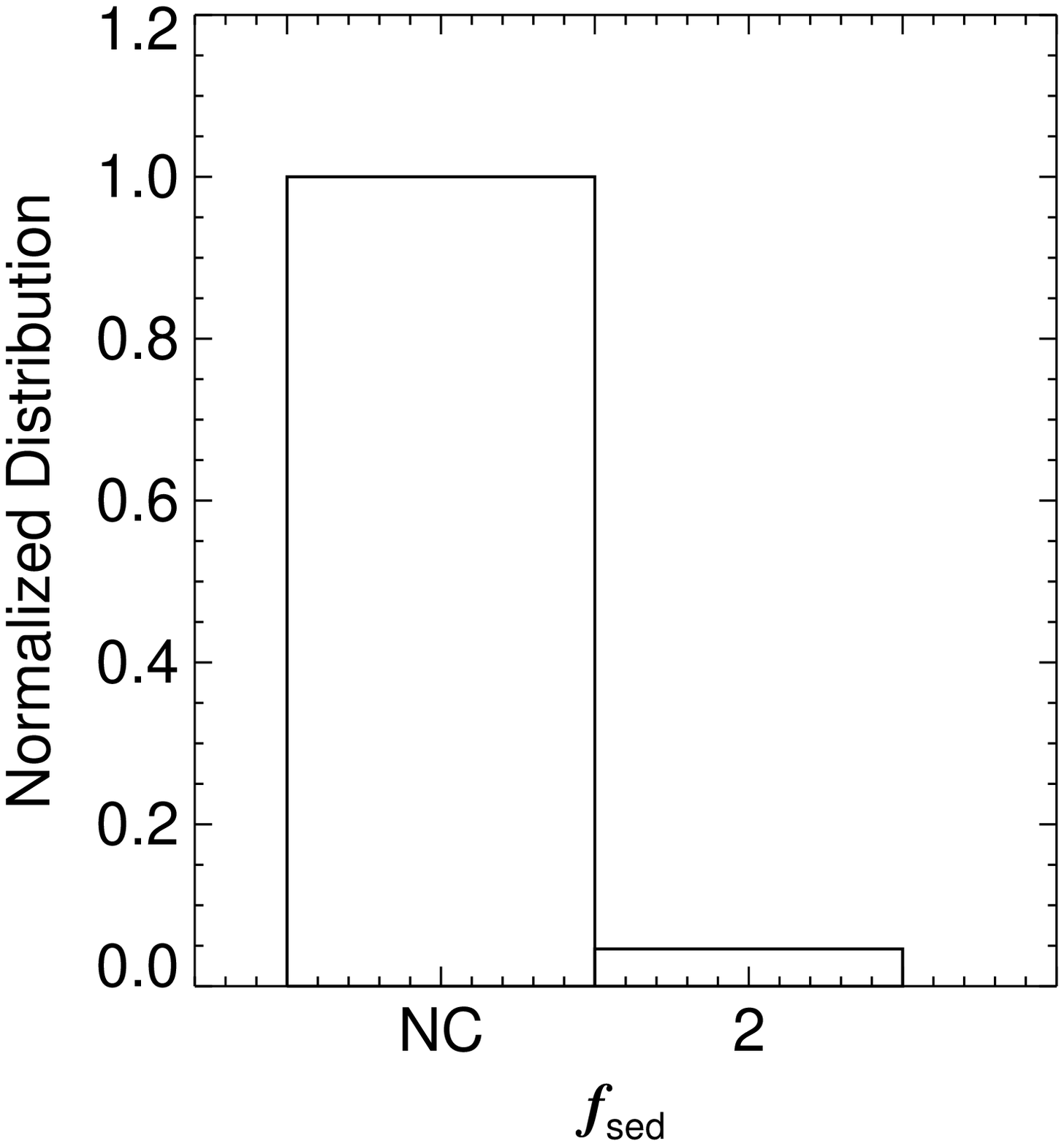}
\caption{Same as Figure~\ref{fig_fit_1617} for WISE~J1812+2721.
\label{fig_fit_1812}}
\end{figure*}

\begin{figure*}
\centering
\epsscale{1.0}
\includegraphics[width=0.9\textwidth]{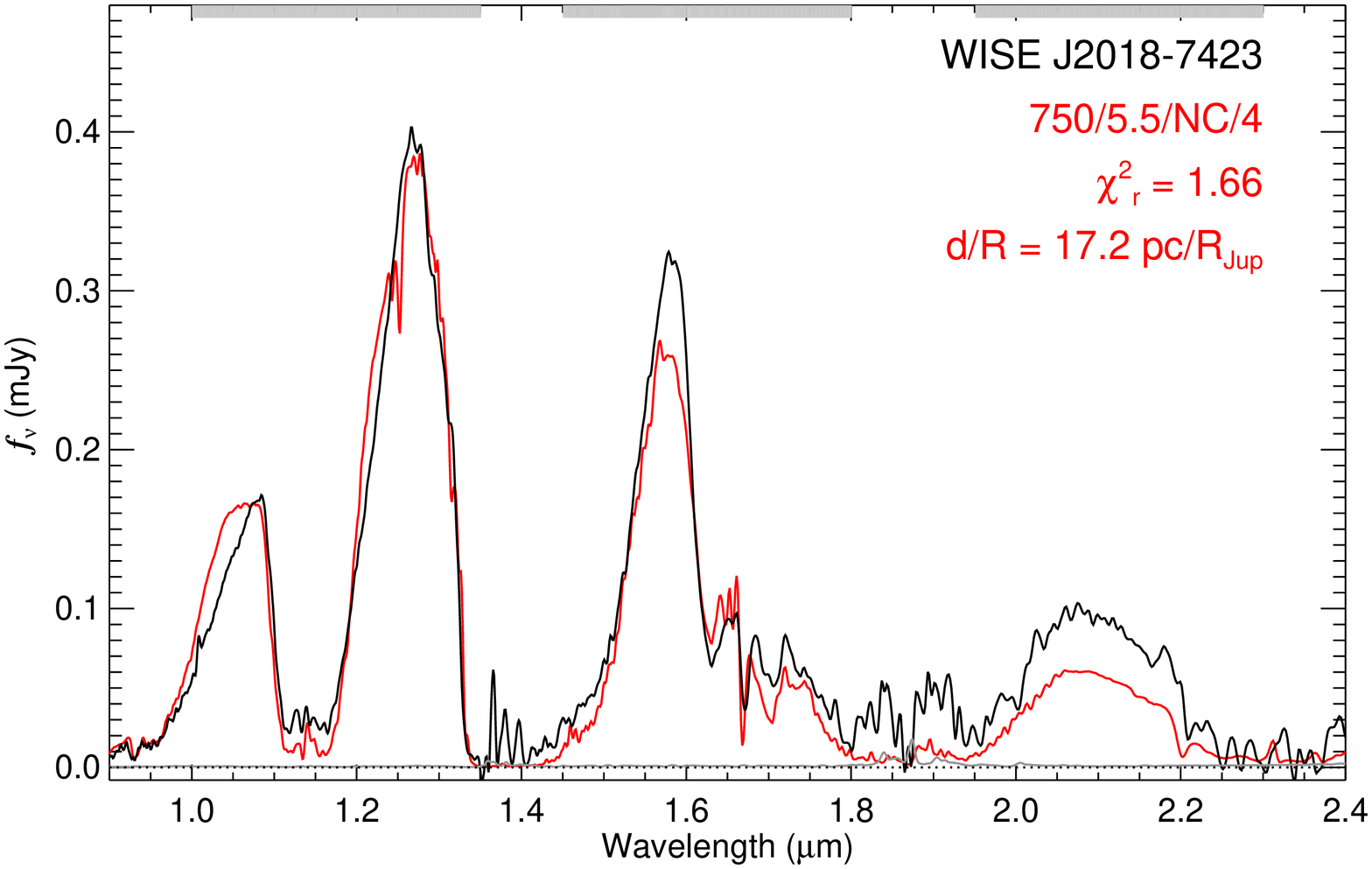}
\includegraphics[width=0.3\textwidth]{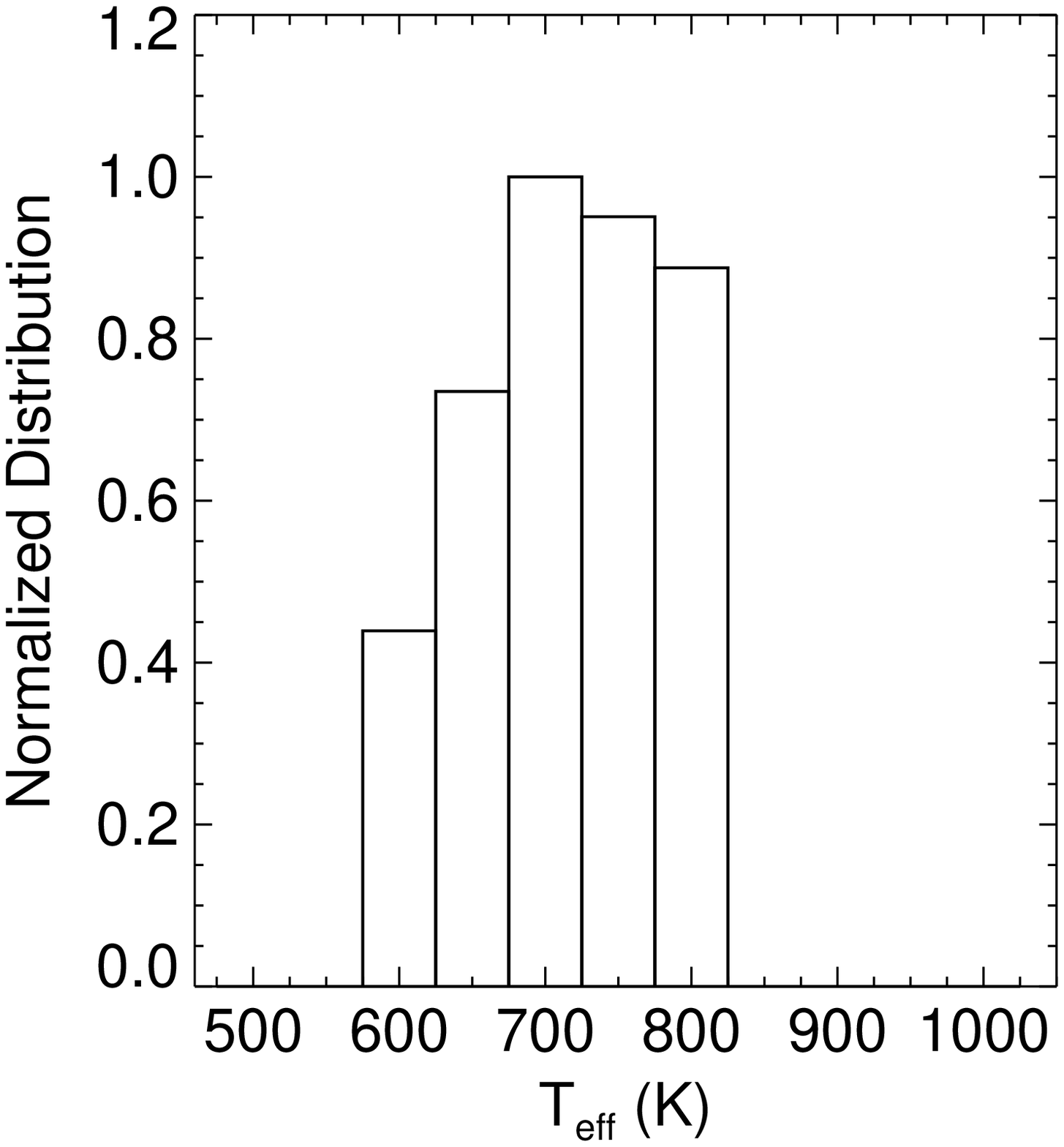}
\includegraphics[width=0.3\textwidth]{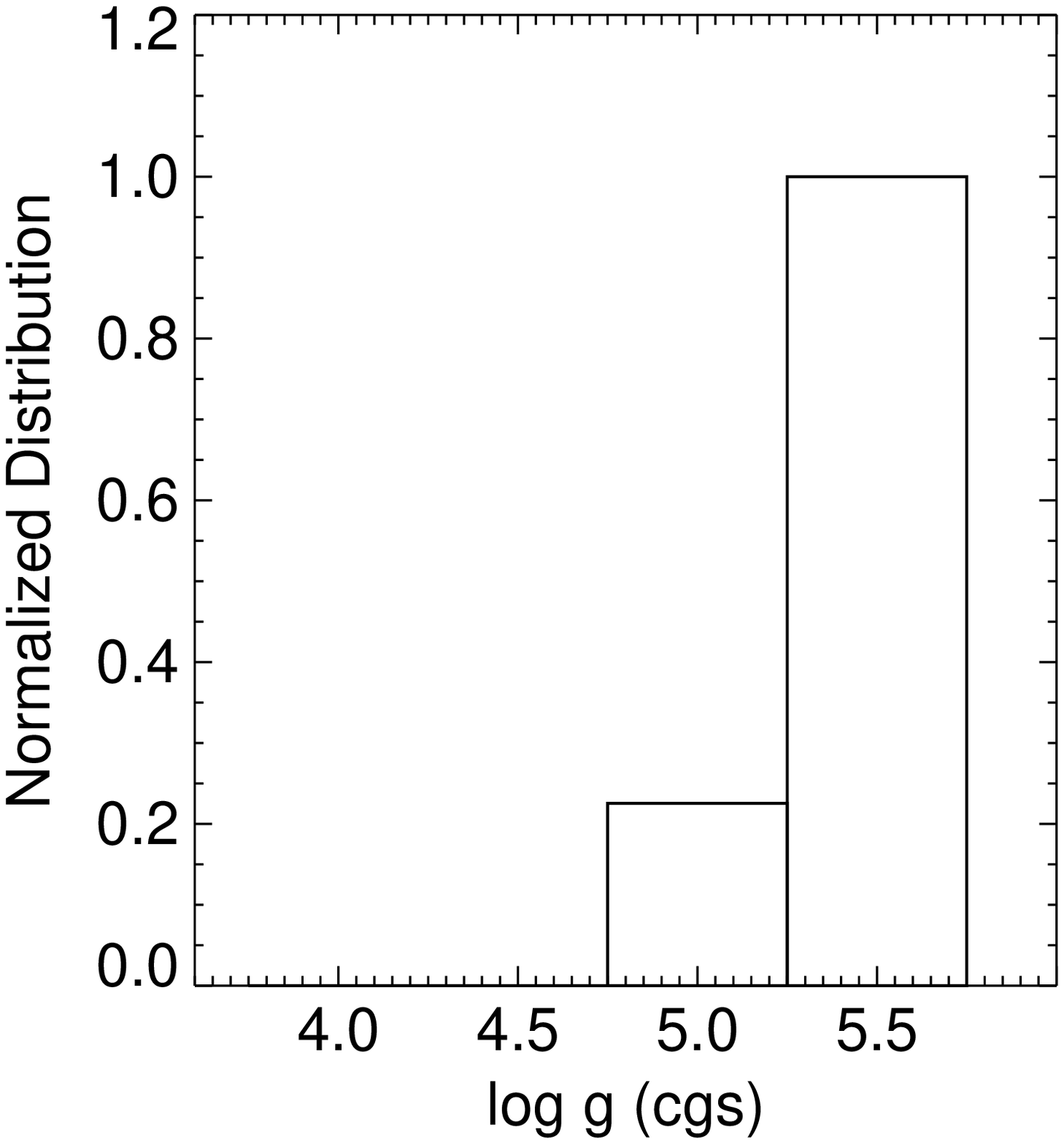}
\includegraphics[width=0.3\textwidth]{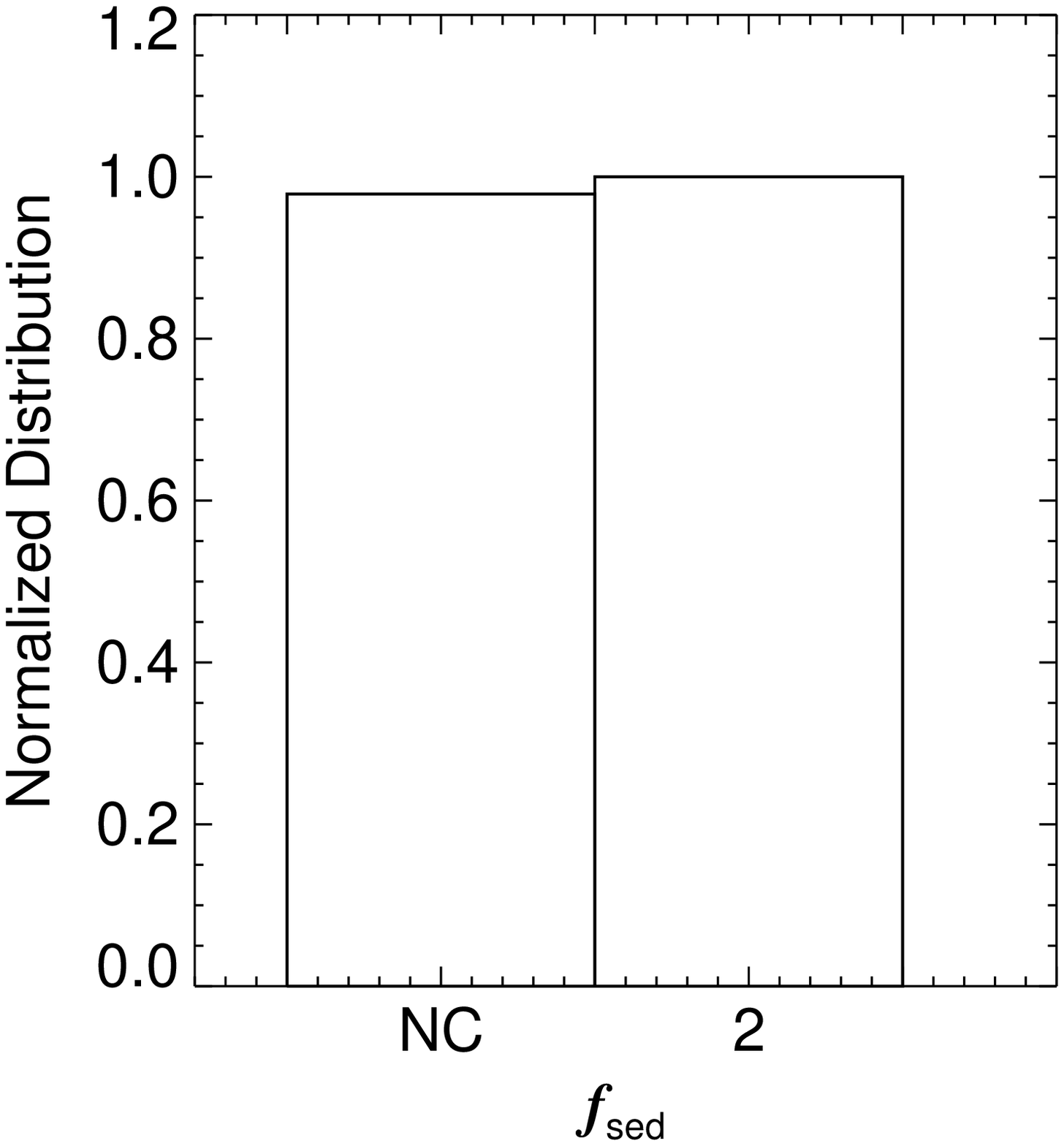}
\caption{Same as Figure~\ref{fig_fit_1617} for WISE~J2018$-$7423.
\label{fig_fit_2018}}
\end{figure*}

\begin{figure*}
\centering
\epsscale{1.0}
\includegraphics[width=0.9\textwidth]{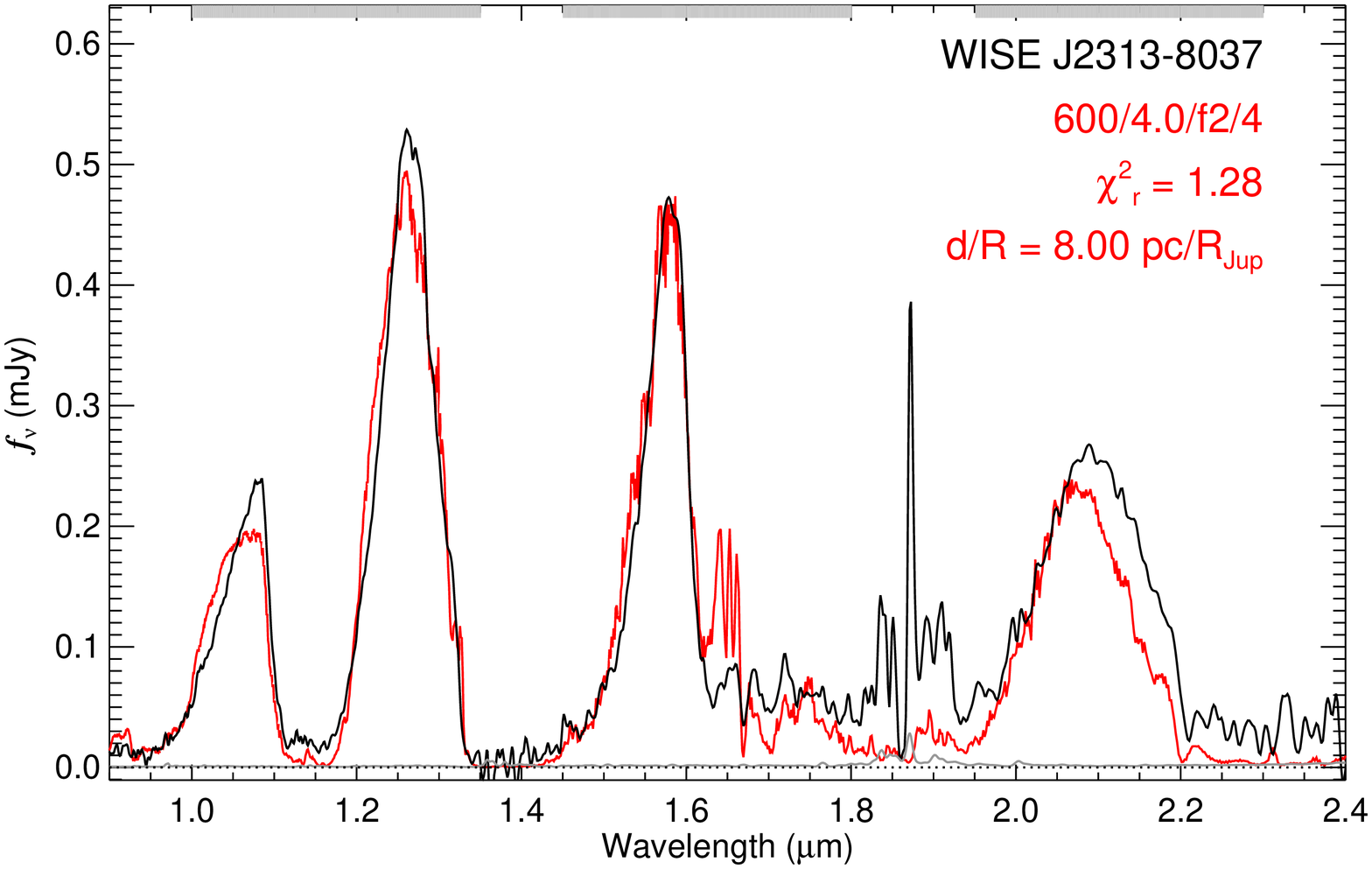}
\includegraphics[width=0.3\textwidth]{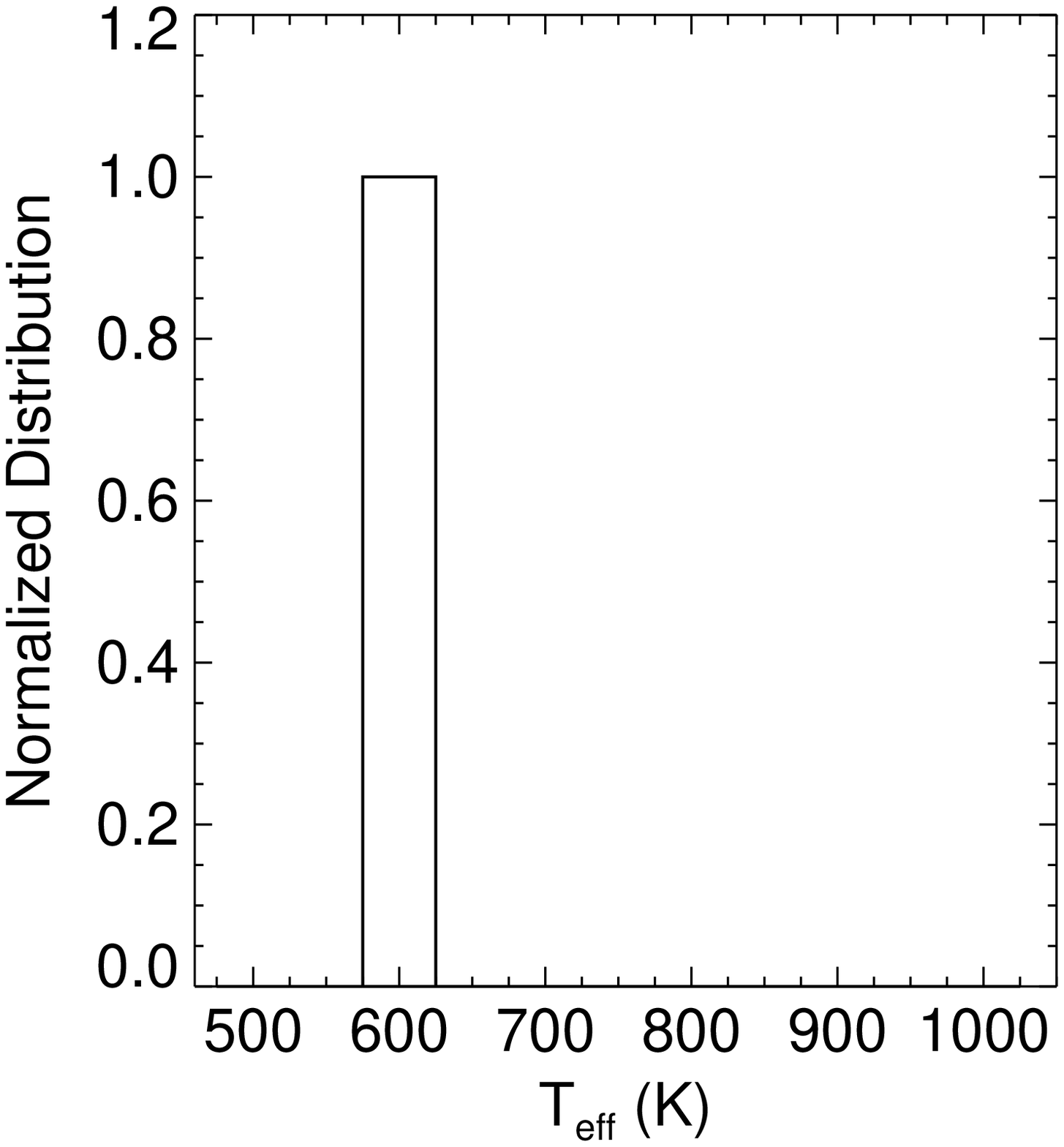}
\includegraphics[width=0.3\textwidth]{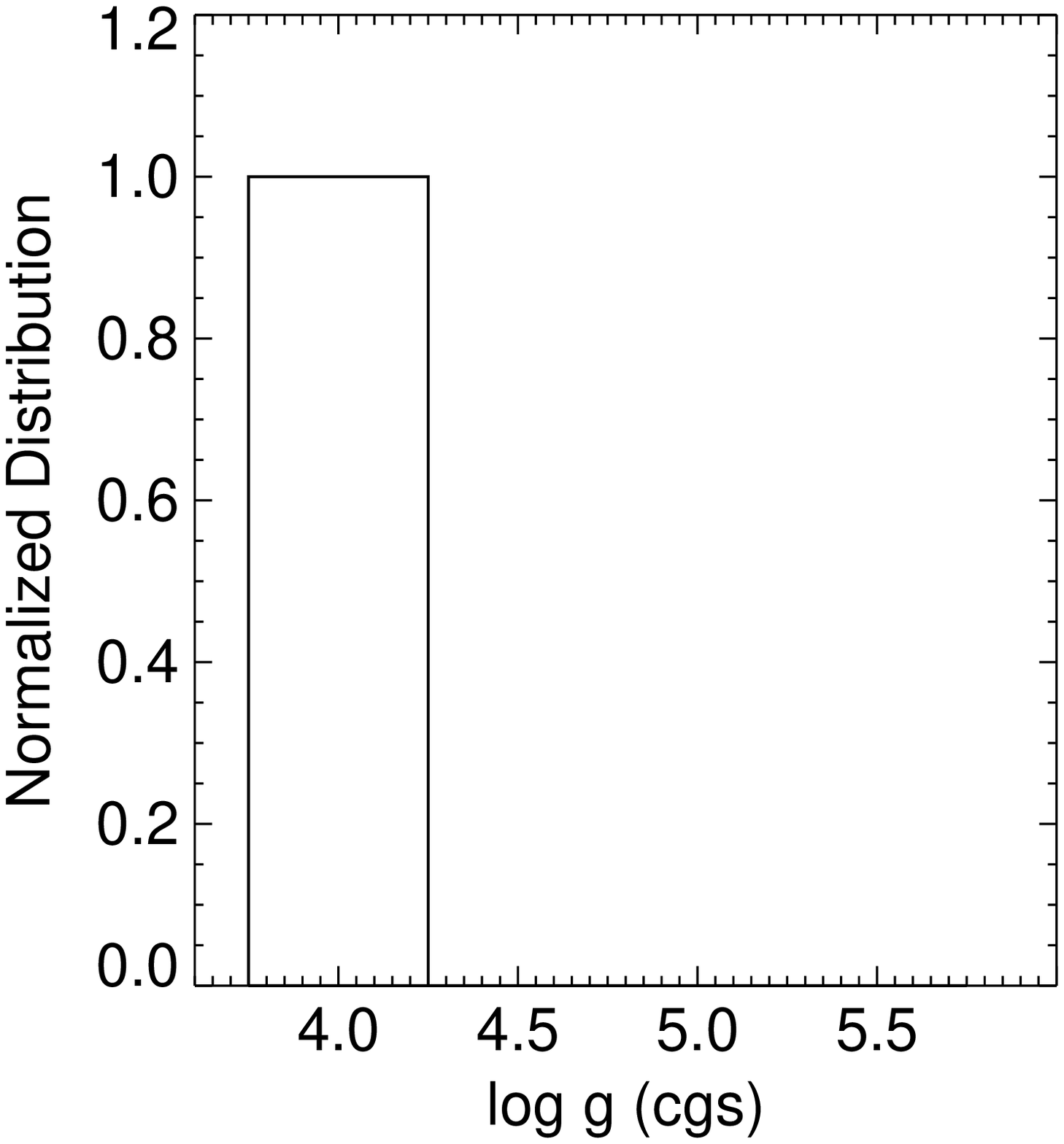}
\includegraphics[width=0.3\textwidth]{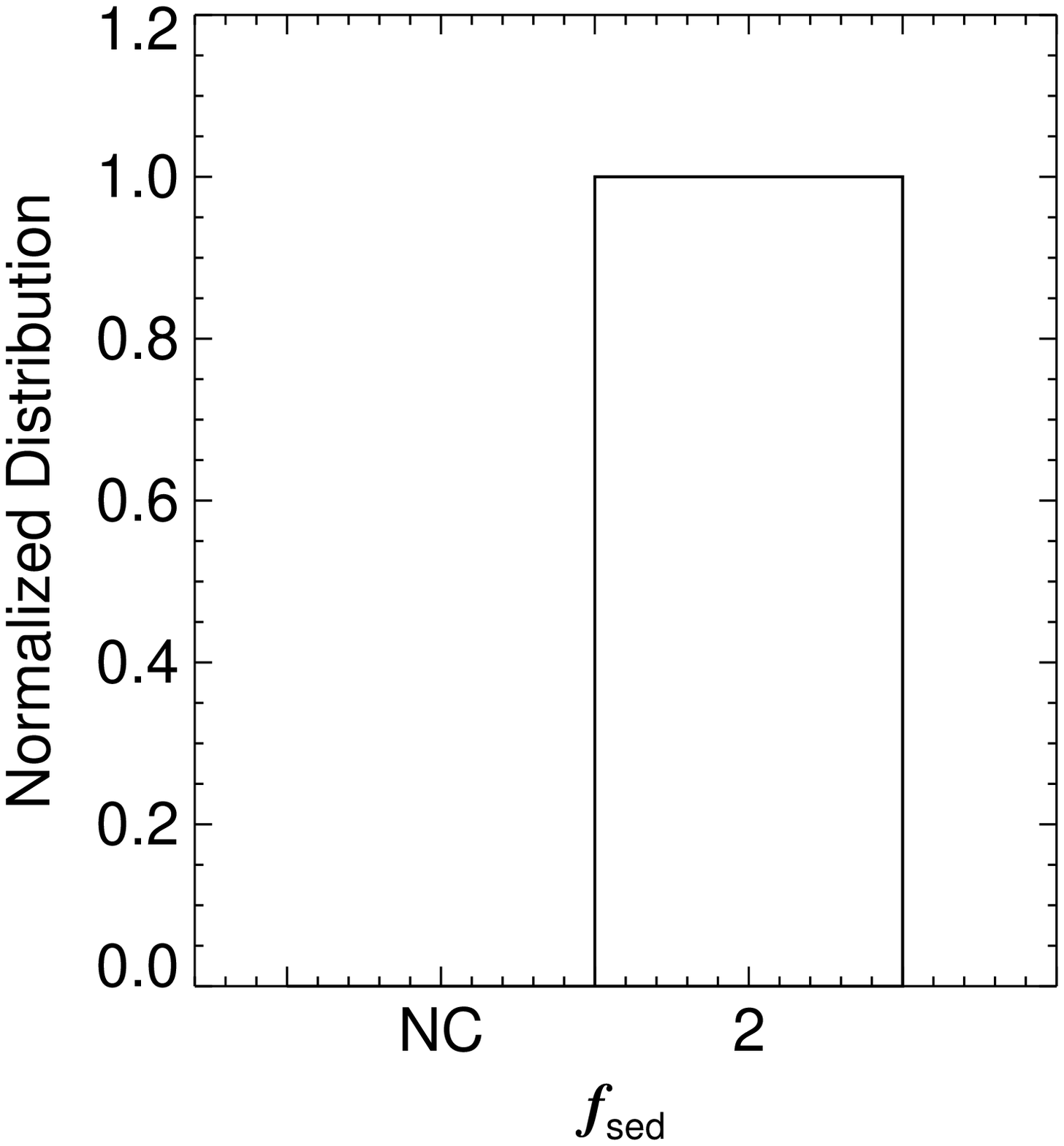}
\caption{Same as Figure~\ref{fig_fit_1617} for WISE~J2313$-$8037.
\label{fig_fit_2313}}
\end{figure*}

\begin{figure*}
\centering
\epsscale{1.0}
\includegraphics[width=0.9\textwidth]{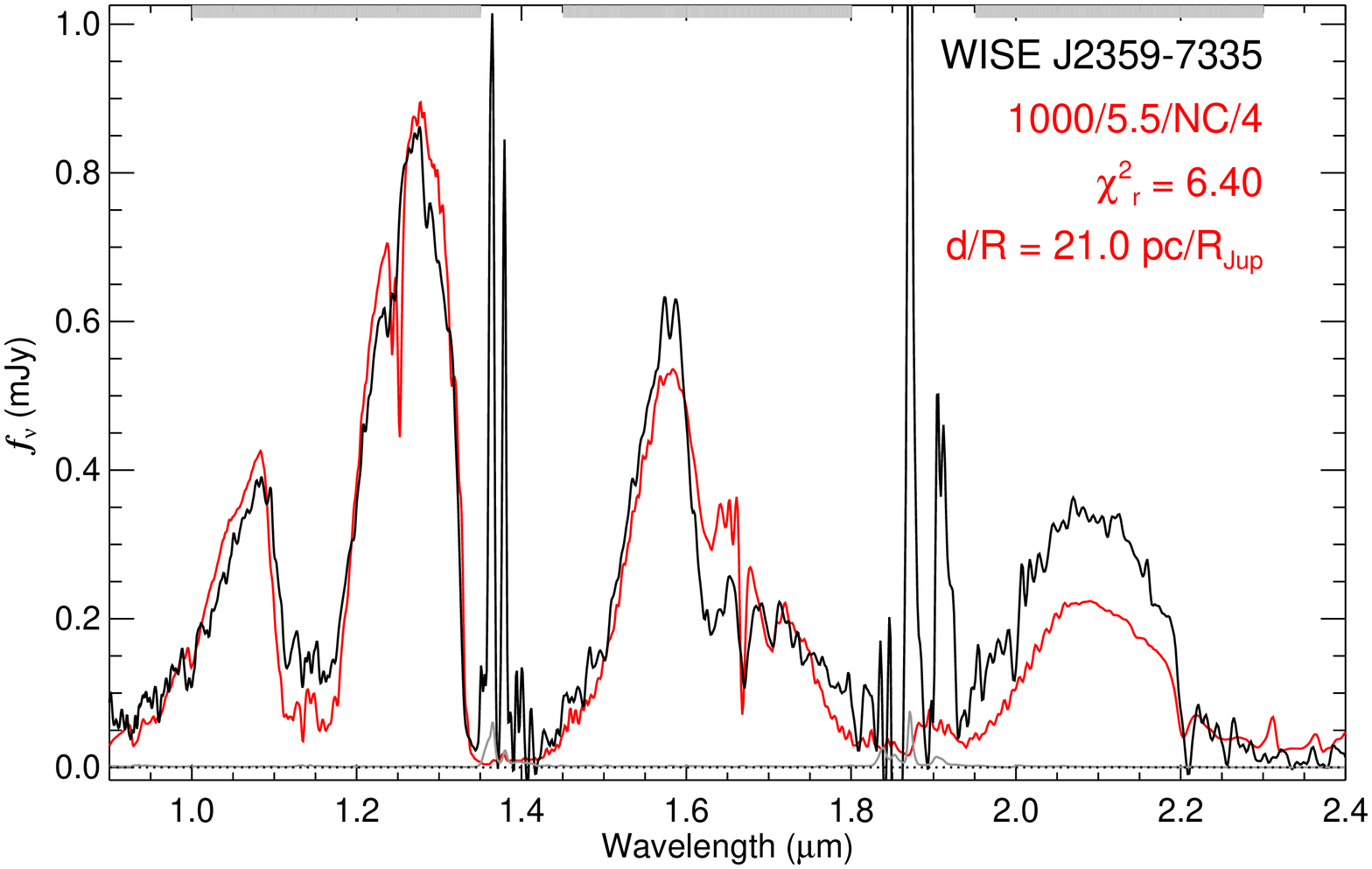}
\includegraphics[width=0.3\textwidth]{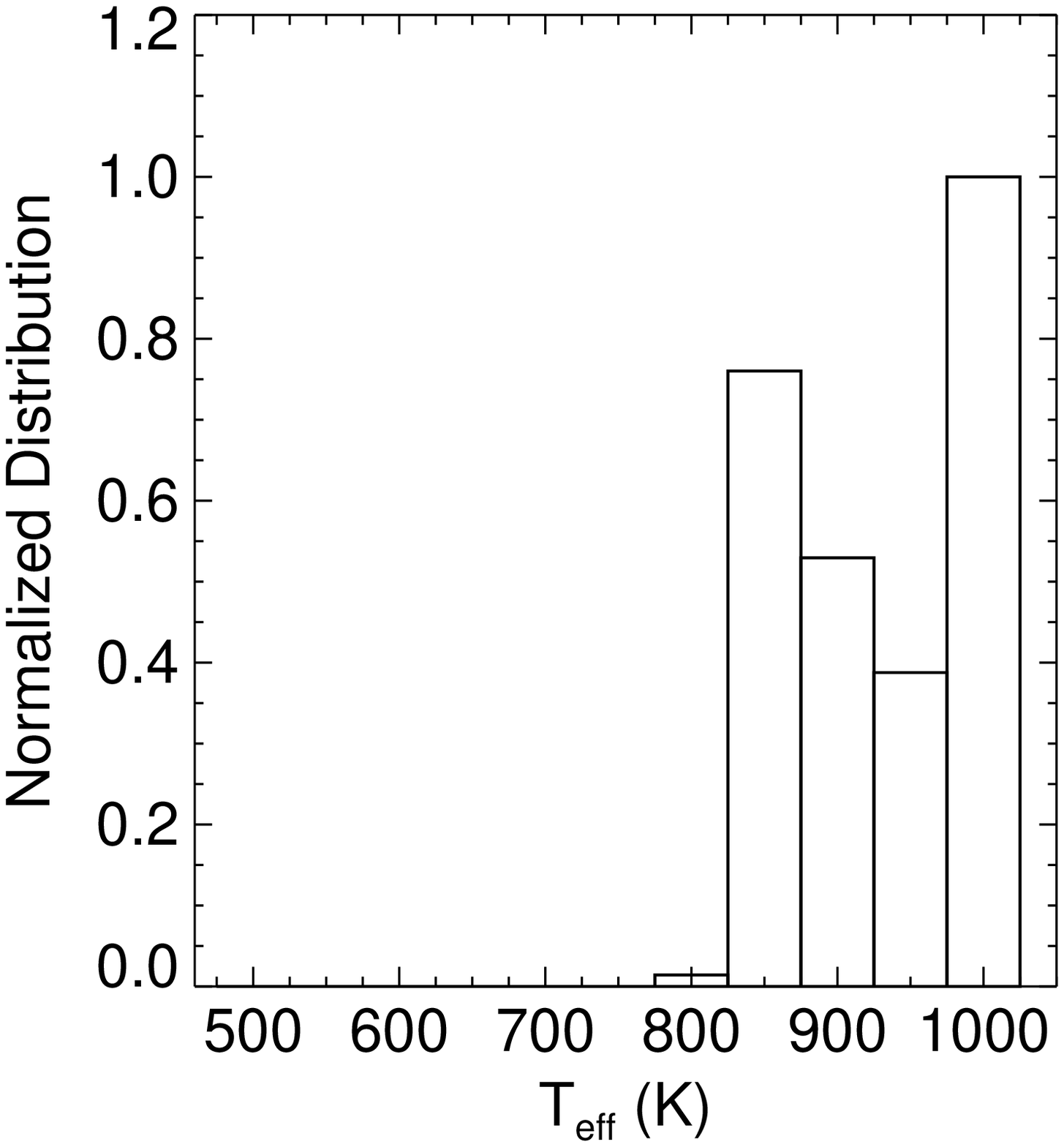}
\includegraphics[width=0.3\textwidth]{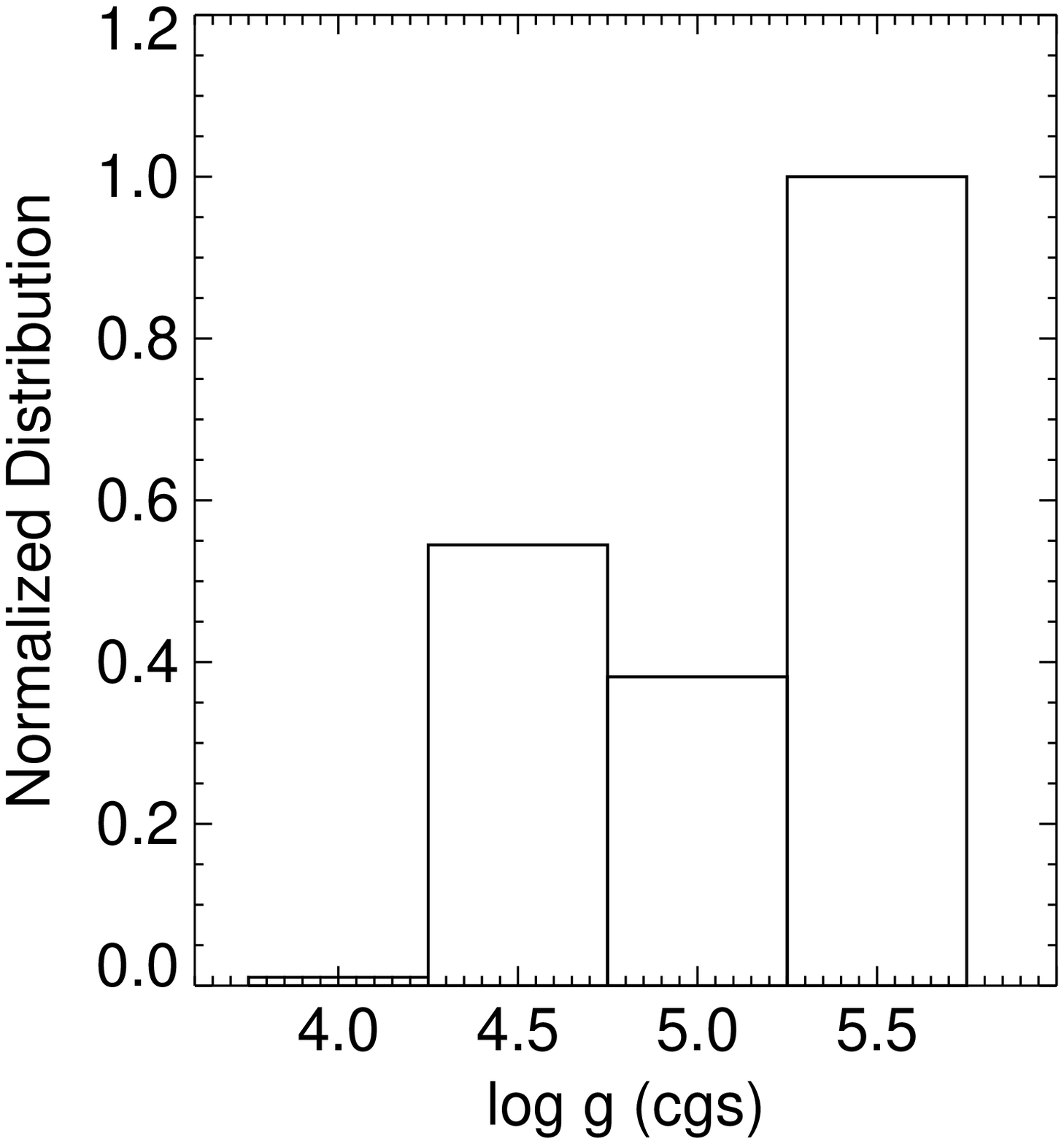}
\includegraphics[width=0.3\textwidth]{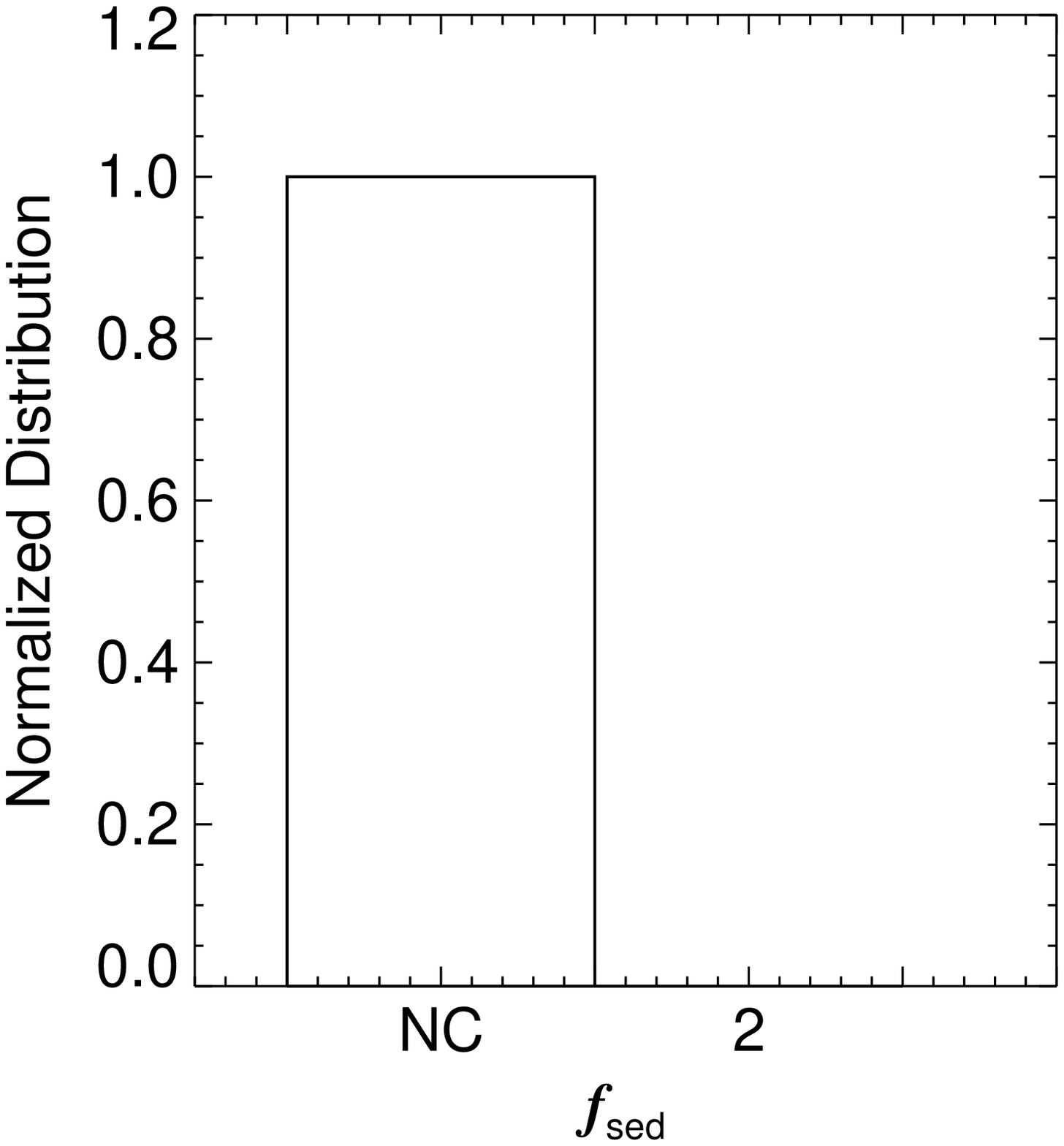}
\caption{Same as Figure~\ref{fig_fit_1617} for WISE~J2359$-$7335.
\label{fig_fit_2359}}
\end{figure*}

\begin{figure*}
\centering
\epsscale{1.0}
\includegraphics[width=0.9\textwidth]{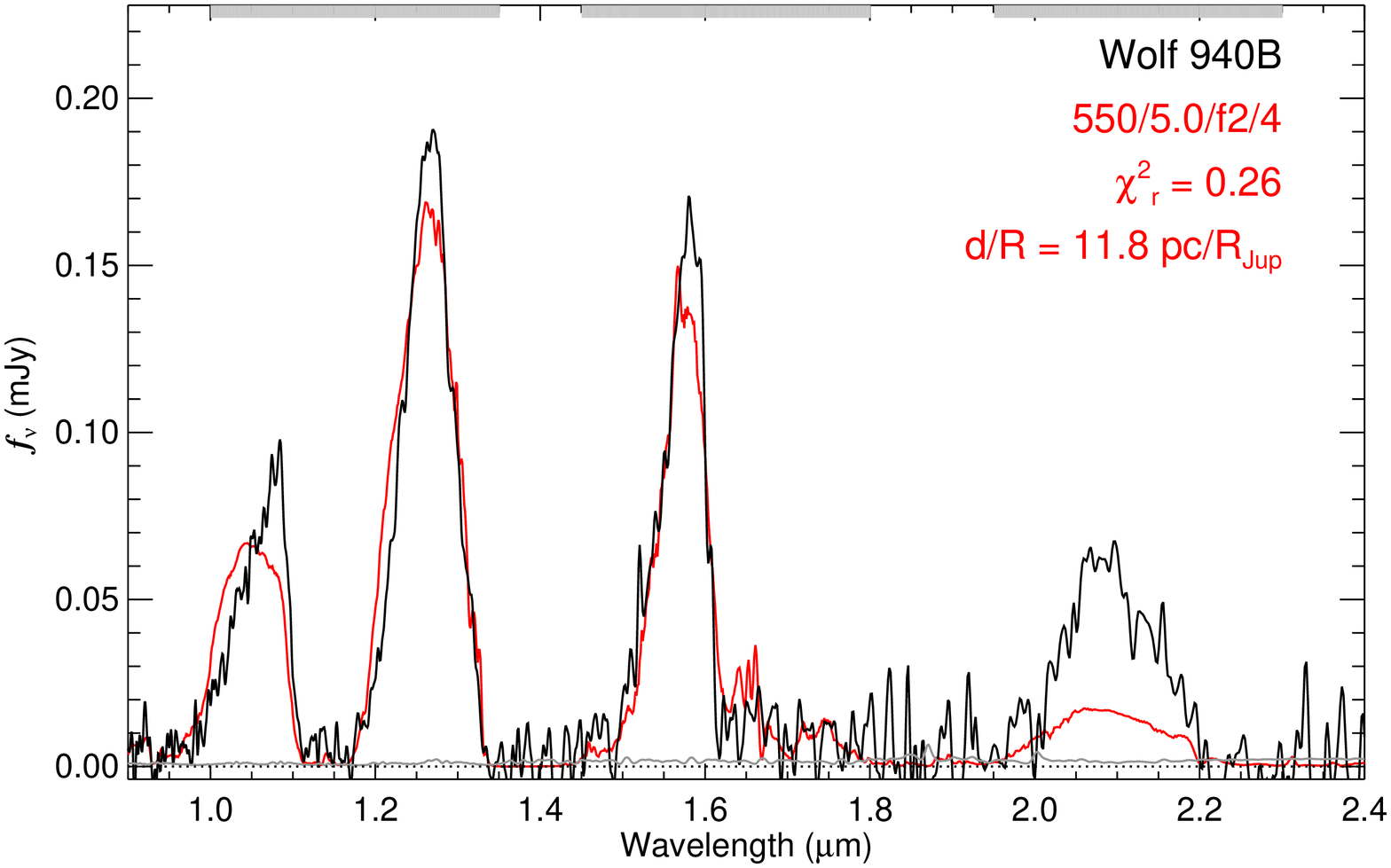}
\includegraphics[width=0.3\textwidth]{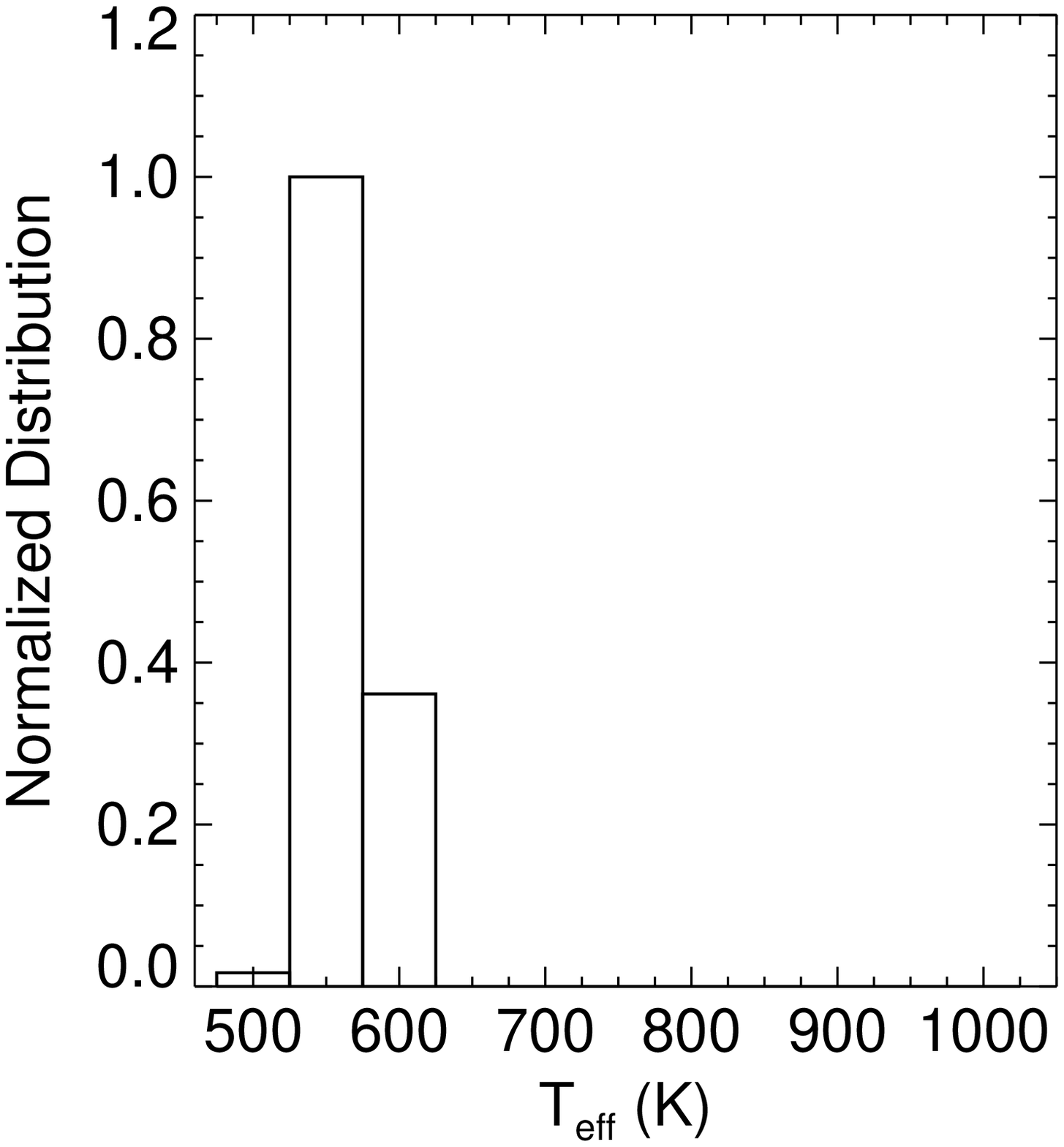}
\includegraphics[width=0.3\textwidth]{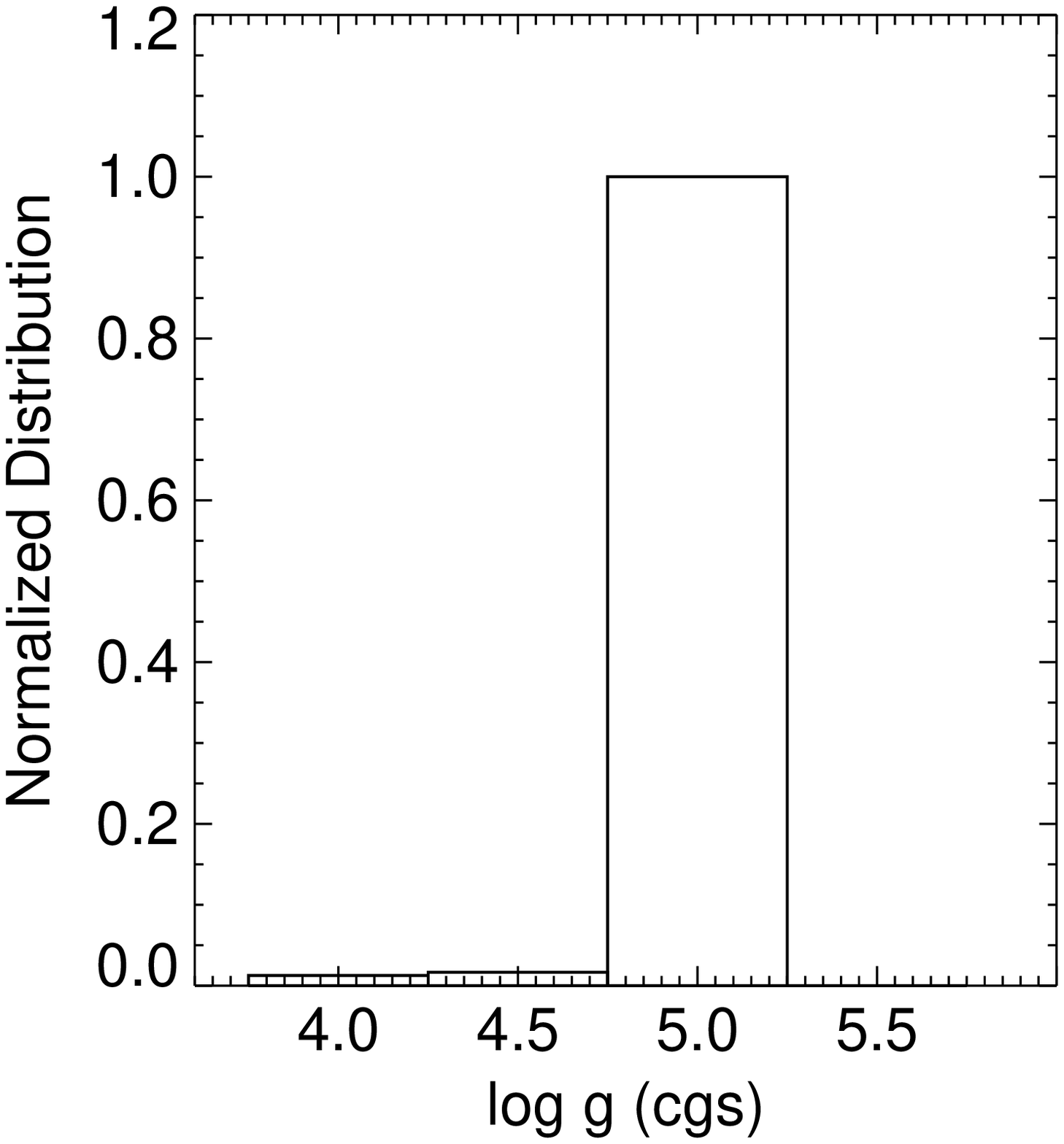}
\includegraphics[width=0.3\textwidth]{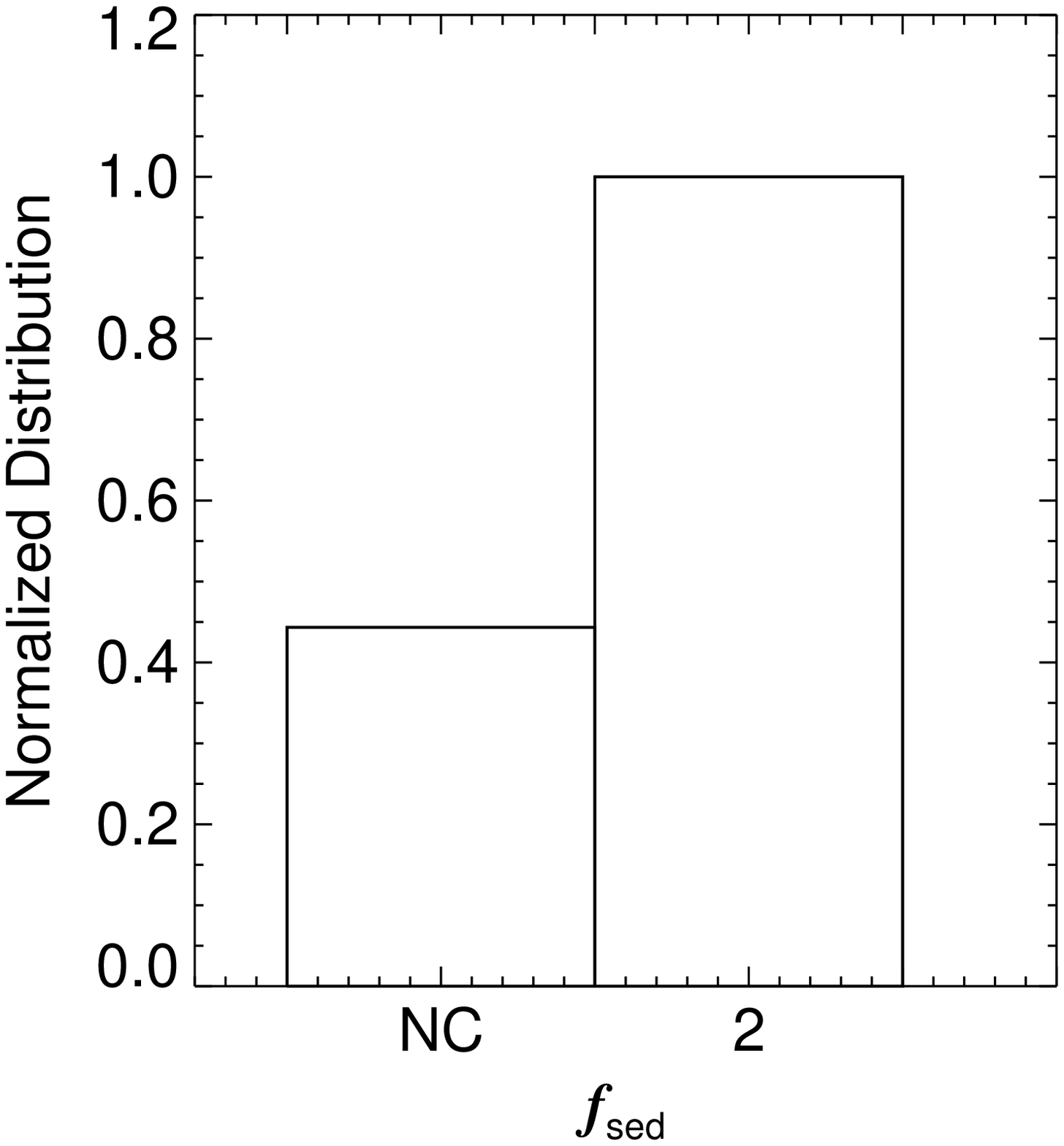}
\caption{Same as Figure~\ref{fig_fit_1617} for Wolf~940B.
\label{fig_fit_wolf940b}}
\end{figure*}

\begin{deluxetable*}{llccccccc}
\tabletypesize{\footnotesize}
\tablecaption{Results from Model Fits to T Dwarf FIRE Spectra. \label{tab_characteristics}}
\tablewidth{0pt}
\tablehead{
 & & & & & & & \colhead{Model Fit} & \colhead{Photometric}  \\
\colhead{Source} &
\colhead{SpT} & 
\colhead{\teff} & 
\colhead{\logg} & 
\colhead{Cloudy?} & 
\colhead{Mass} & 
\colhead{Age} &
\colhead{Distance} & 
\colhead{Distance} \\ 
\colhead{} &
\colhead{} & 
\colhead{(K)} & 
\colhead{({\cmss})} & &
\colhead{({\mjup})} & 
\colhead{(Gyr)} &
\colhead{(pc)} & 
\colhead{(pc)}  \\
}
\startdata
WISE J1617+1807 & T8 & 600$^{+30}_{-30}$ &  4.0$^{+0.3}_{-0.3}$ & Yes  & 7$\pm$3 &  0.2$\pm$0.3 &  13.1$\pm$0.6 & 13.0$\pm$1.5  \\
WISE J1812+2721 & T8.5: & 620$^{+30}_{-30}$ &  4.3$^{+0.3}_{-0.3}$ & No   & 13$\pm$7 &  0.9$\pm$1.3 &  19$\pm$3 & 13$\pm$3  \\
WISE J2018-7423 & T7 & 710$^{+50}_{-60}$ &  5.4$^{+0.3}_{-0.3}$ & Yes?  & 50$\pm$9 &  10$\pm$4 &  12.2$\pm$2.3 & 13.1$\pm$1.4  \\
WISE J2313-8037 & T8 & 600$^{+30}_{-30}$ &  4.0$^{+0.3}_{-0.3}$ & Yes  & 7$\pm$3 &  0.3$\pm$0.4 &  9.3$\pm$0.4 & 11.7$\pm$1.6  \\
WISE J2359-7335 & T5.5 & 930$^{+50}_{-50}$ &  5.1$^{+0.4}_{-0.4}$ & No  & 38$\pm$18 &  4$\pm$4 &  17$\pm$3 & 12.5$\pm$1.7  \\
Wolf 940B & T8.5 & 560$^{+30}_{-30}$ &  5.0$^{+0.3}_{-0.3}$ &  Yes? & 30$\pm$10 &  7$\pm$4 &  11.5$\pm$1.6 & 12.5$\pm$0.7\tablenotemark{a}  \\
\enddata
\tablenotetext{a}{Parallax distance measurement for the Wolf~940A primary \citep{1980AJ.....85..454H}.} 
\end{deluxetable*}

We followed a fitting prescription similar to that described in \citet{2010ApJ...725.1405B},
built upon contemporary work by \citet{2008ApJ...678.1372C} and \citet{2009ApJ...706.1114B}. 
We used solar metallicity models with non-equilibrium chemistry 
(eddy diffusion parameter {\kzz} = 10$^4$~{\cmss}; \citealt{1999ApJ...519L..85G, 2006ApJ...647..552S, 2007ApJ...669.1248H}),
and considered both cloud-free and cloudy models, the latter with condensate sedimentation parameter
{\fsed} = 2 \citep{2001ApJ...556..872A}.
Atmospheric parameters {\teff} = 500--1000~K (50~K steps) and {\logg} = 4.0--5.5~cgs (0.5~dex steps) were sampled, with corresponding physical parameters (mass, age and radius) determined using the appropriate evolutionary tracks from \citet{2008ApJ...689.1327S}.
The FIRE spectra were scaled to the apparent $J$-band magnitude of each source,
and both models and data were smoothed to a common resolution of {\ldl} = 300 
and sampled at 4 pixels per resolution element to match FIRE's projected slit width.
Spectra were compared in the 1.0--1.35~$\micron$, 1.45--1.8~$\micron$ and 1.95--2.3~$\micron$
regions, using the $\chi^2$ statistic to assess both the goodness-of-fit and the relative scaling factor $C \equiv (R/d)^2$, where $R$ is the radius of the brown dwarf and $d$ its distance from the Sun.   
We further constrained our fits by requiring that the model-inferred distance be within 3$\sigma$ of the estimated distance based on $W2$ photometry (Section~4.2).  Note that this constraint is only weakly sensitive to unresolved multiplicity since both distances are based on photometric scaling.
Means and uncertainties in the atmospheric parameters were determined using the F-distribution probability distribution function (F-PDF) as a weighting factor (Equations~1--4 in \citealt{2010ApJ...725.1405B}).
Sampling uncertainties of 25~K and 0.25~dex were also imposed on the inferred {\teff} and {\logg} values, 
which were propagated into the estimated physical parameters.

Figures~\ref{fig_fit_1617}--\ref{fig_fit_wolf940b} show the best-fitting models for each of the WISE spectra and for Wolf~940B, as well as the F-PDF weighted distributions of {\teff}, {\logg} and {\fsed} parameters.
Table~\ref{tab_characteristics} summarizes the inferred atmospheric and physical parameters.
Overall, the models provide reasonable fits to the spectral data, with the exception of known discrepancies
in the core of the 1.6--1.7~$\micron$ {\meth} band, the strength of the 1.25~$\micron$ K~I lines (for the warmer T dwarfs), and the detailed shape of the 1.05~$\micron$ $Y$-band peak.
Fits to WISE~J2359$-$7335 are particularly poor, likely due to the best fitting models residing at the end of our parameter range.  For Wolf~940B, the best-fitting models poorly reproduce the brightness of the
observed $K$-band peak.  Examining the inferred parameters in detail, we find that {\teff}s track well with spectral type and are consistent with the spectral type/{\teff} scales of \citet{2004AJ....127.3516G, 2009ApJ...702..154S} and \citet{2010A&A...524A..38M}.  This correlation may be an artifact of the imposed distance constraints, which are tied to the $M_{W2}$/spectral type relation defined above.  However, our inferred parameters for Wolf~940B, which are constrained by the parallactic distance of the system \citep{1980AJ.....85..454H}, are consistent with the broad-band spectral fitting results of \citet{2010ApJ...720..252L}.  We infer similar {\teff}s for the WISE targets when the distance constraint is removed.  Surprisingly, a range of cloud parameters are indicated, with both WISE~J1617+1807 and WISE~J2313$-$8037 exhibiting evidence for the presence of
photospheric cloud opacity.  We discuss some of these secondary parameters in further detail below.

\section{Discussion of Individual Sources}

\subsection{The T8.5: Dwarf WISE~J1812+2721}

\begin{figure}
\centering
\epsscale{0.9}
\plotone{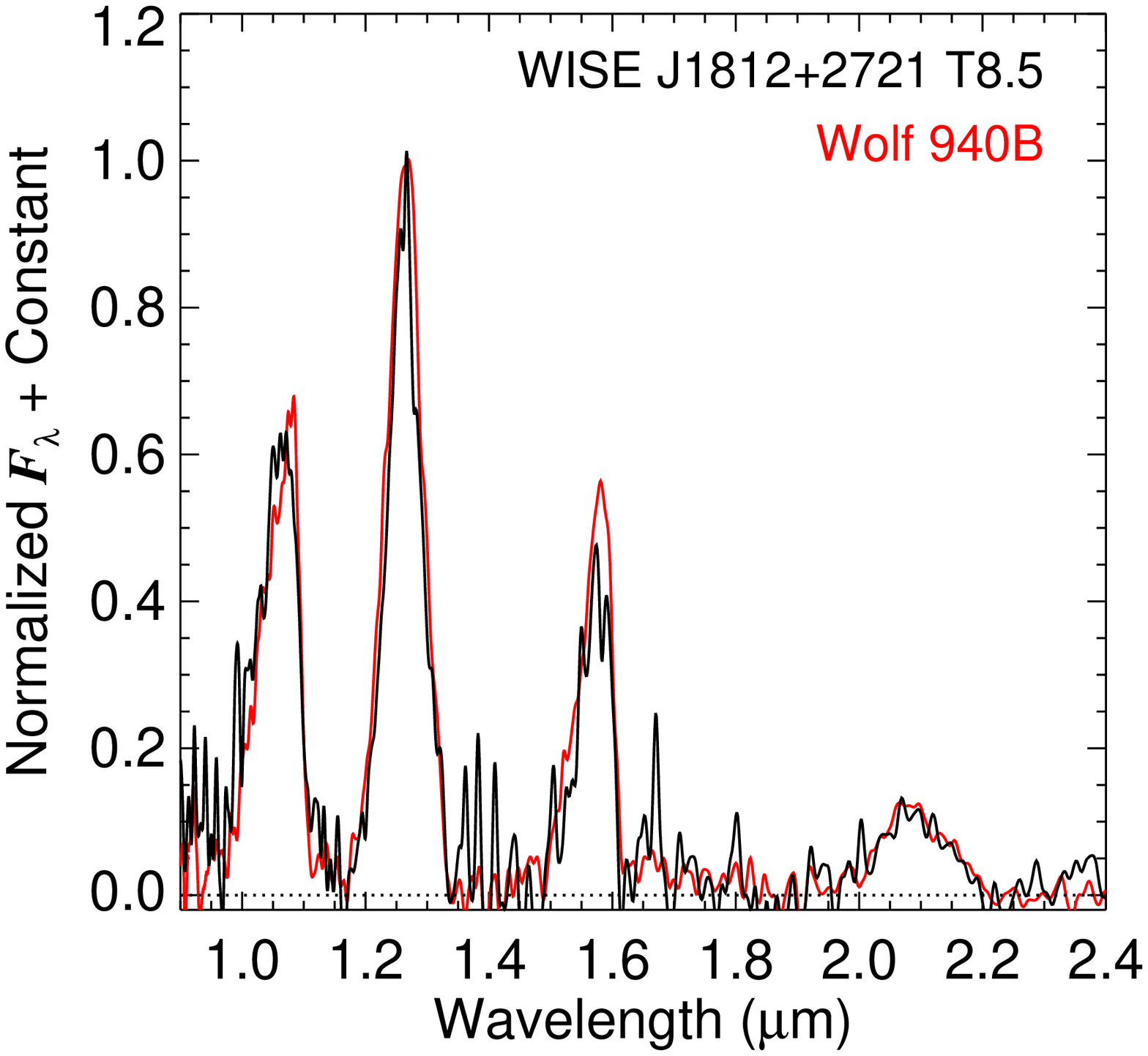}
\caption{Comparison of FIRE spectra for WISE~J1812+2721 (black line)
and Wolf~940B (red line), both smoothed to a resolution of {\ldl} = 150
and normalized at 1.27~$\micron$.
\label{fig_comp1812}}
\end{figure}

The latest-type source in this sample is WISE~J1812+2721, tentatively classified T8.5: based on spectral comparison to 2MASS~J0415-0935 in Figure~\ref{fig_class} and spectral indices.  
It also has the reddest $W1-W2$ color in our sample.  
While this classification is somewhat uncertain due to the lower signal-to-noise of the spectral data,
its binned spectrum is very similar to that of Wolf~940B, itself classified T8.5 (Figure~\ref{fig_comp1812}).
Near- and mid-infrared spectral model fits to Wolf~940B indicate {\teff} = 585--625~K and {\logg} = 4.83--5.22~cgs
for an age of 3--10~Gyr \citep{2010ApJ...720..252L}, similar to the results we infer here.
For WISE~J1812+2721 we find a comparable {\teff} but much lower surface gravity, age and mass.  
While surface gravity determinations should in general be treated with caution (see below),
the agreement in spectral morphology and inferred {\teff} confirms the cool nature of this source.

\subsection{WISE~J1617+1807 and WISE~J2313$-$8037: Young and Cloudy Field T Dwarfs?}

WISE~J1617+1807 and WISE~J2313$-$8037 exhibit relatively red $J-K$ spectrophotometric colors and large $K/J$ ratios for their spectral types, and our spectral model fits suggest cool ({\teff} = 600~K), low surface gravity ({\logg} = 4.0~cgs), and cloudy atmospheres.
The inferred surface gravities are driven largely by the relatively bright $K$-band peaks of these spectra, an indication of reduced collision induced {\hh} absorption; surface gravity variations in the absorption strength of this molecule is also cited as an explanation for the red near-infrared colors of young, low-mass L dwarfs
(e.g., \citealt{2001MNRAS.326..695L, 2006ApJ...639.1120K, 2008ApJ...689.1295K, 2007ApJ...657..511A}).
Thick clouds may also give rise to reddened $J-K$ colors in L dwarf spectra (e.g., \citealt{2004AJ....127.3553K, 2009ApJ...702..154S}), so it is pertinent that the spectra of both WISE~J1617+1807 and WISE~J2313$-$8037 are best fit by cloudy models.  Cloud opacity primarily influences the
$YJH$ flux peaks in brown dwarf spectra, which represent minima in gas opacity \citep{2001ApJ...556..872A}.  As such, the models without clouds exhibit $J$-band peaks that are too strong for these two sources.  This discrepancy drives our model fits toward cloudier atmospheres.
        
The presence of clouds has recently been suggested in similar model fits to the T8 dwarf Ross~458C, a widely-separated companion to a nearby M dwarf binary system which has an independent age
constraint of 150--800~Myr \citep{2010ApJ...725.1405B, 2010MNRAS.405.1140G, 2010A&A...515A..92S}. 
The similarity in the inferred properties of Ross~458C,  WISE~J1617+1807 and WISE~J2313$-$8037---low temperature, low surface gravity and cloudy atmospheres---appears indicative of a trend toward cloudier atmospheres in younger brown dwarfs.  Indeed, such a trend has 
previously been proposed to explain the spectra of young L dwarfs (e.g., \citealt{2006ApJ...651.1166M, 2009ApJ...702..154S}) and in contemporary studies of directly-detected exoplanets (e.g., \citealt{2010ApJ...723..850B, 2011ApJ...729..128C, 2011arXiv1102.5089M}).
While compelling, evidence for these trends are not yet conclusive.  
Our model fits for the WISE T dwarfs indicate relatively
young ages ($\sim$200--300~Myr) and low masses ($\sim$7~{\mjup}), values that are somewhat suspect for a pair of isolated field objects (although we cannot rule out membership
in a nearby young association such as AB Doradus or Tucana Horologium; \citealt{2001ApJ...559..388Z, 2004ApJ...613L..65Z}).
The fits are also constrained by fairly uncertain spectrophotometric distance estimates.
Moreover, we have not considered metallicity variations in this study which are also known to modulate the $K$-band peaks of both L and T dwarf spectra (\citealt{2006ApJ...639.1095B, 2007ApJ...658..617B, 2007ApJ...660.1507L, 2008ApJ...686..528L}).  
We therefore regard the increased role of clouds in shaping young T dwarf spectra as a suggestive trend,
and defer further analysis to more comprehensive, broad-band spectral modeling (M.\ Cushing, in prep.).

\subsection{WISEPC~J201824.98$-$742326.1: An Old Blue T Dwarf?}

WISE~J2018$-$7423 exhibits an opposing spectral peculiarity: a
suppressed $K$-band peak resulting in an unusually blue 
spectrophotometric near-infrared color ($J-K$ = $-$0.54$\pm$0.10) 
and small $K/J$ index (0.097$\pm$0.003) for its spectral type. 
Previously identified blue T dwarfs, 
such as 2MASS J09373487+2931409 ($J-K$ = $-$1.10$\pm$0.07; $K/J$ = 0.08; \citealt{2002ApJ...564..421B, 2006ApJ...639.1095B, 2004AJ....127.3553K})
and SDSS~J141624.08+134826.7B (hereafter SDSS~J1416+1348B, $J-K$ = $-$1.58$\pm$0.17; $K/J$ = 0.037$\pm$0.004; \citealt{2010MNRAS.404.1952B, 2010AJ....139.2448B, 2010A&A...510L...8S}),
have similarly suppressed $K$-band peaks from
strong collision-induced {\hh} absorption,
attributed to a high surface gravity and/or subsolar metallicity.
Our spectral model fits support a high surface gravity for this source, 
indicating {\logg} $\sim$ 5.4~cgs, age $\tau$ $\gtrsim$ 6~Gyr and mass M $\sim$ 50~{\mjup}; the 
estimated {\vtan}  = 56$\pm$6~{\kms} of this object
supports a relatively old age.
However, our fits cannot test whether this source is metal-poor.

\begin{figure}
\centering
\epsscale{0.9}
\plotone{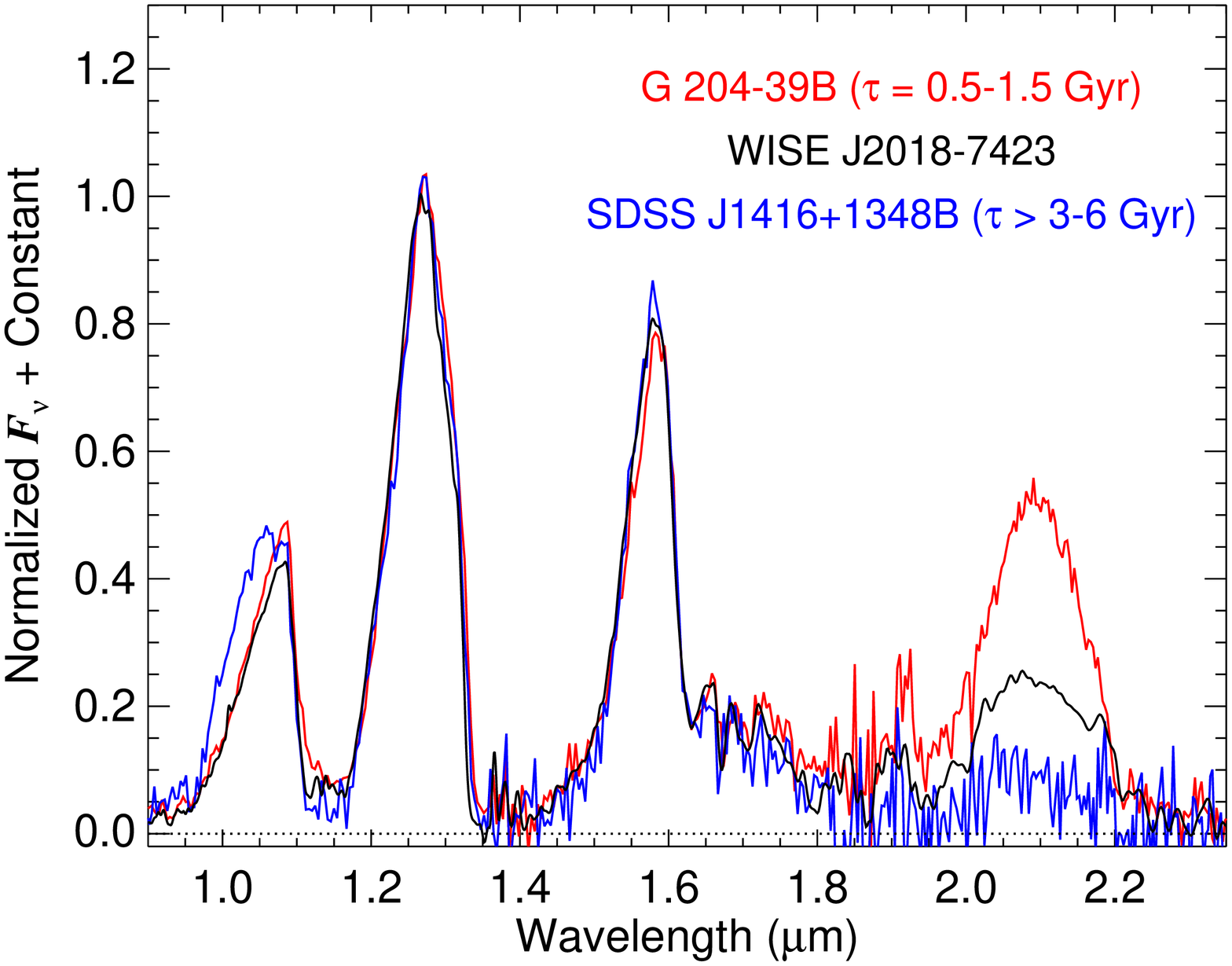}
\caption{Comparison of the near-infrared spectra of WISE~J2018$-$7423 (black line; FIRE data) and the T dwarf companions G~204-39B (red line; SpeX data from \citealt{2006ApJ...639.1095B})
and SDSS~J1416+1348B (blue line; SpeX data from \citealt{2010AJ....139.2448B}).
All three spectra are normalized at their 1.27~$\micron$ spectral peaks.
\label{fig_comp2018}}
\end{figure}

Fortunately, metallicity effects can be separately discerned in the 1.05~$\micron$
$Y$-band peak, which is broadened in both the theoretical and observed spectra
of metal-poor T dwarfs \citep{2006ApJ...639.1095B, 2010AJ....139.2448B}.  
In Figure~\ref{fig_comp2018} we compare the spectrum
of WISE~J2018$-$7423 to those of two
equivalently-classified T dwarf companions to stars with independent age and metallicity constraints: 
the young, metal-rich T6.5 G~204-39B (a.k.a. SDSS~J175805.46+463311.9; \citealt{2004AJ....127.3553K, 2010AJ....139..176F})
and the old, metal-poor T7 SDSS~J1416+1348B.
G~204-39A is an M3 star which exhibits weak signatures of H$\alpha$ and X-ray activity consistent with $\tau$  = 0.5--1.5~Gyr, and optical spectral indicators (i.e., ratio of TiO/CaH) suggesting a slightly supersolar metallicity.
SDSS~J1416+1348A is an unusually blue L dwarf, and 
spectral model fits to both primary and secondary indicate an older ($\tau > 3$~Gyr) and possibly metal-poor system  (e.g., \citealt{2010ApJ...710...45B, 2010AJ....139.2448B, 2010ApJ...725.1405B, 2010AJ....140.1428C}.  
As Figure~\ref{fig_comp2018} shows, the spectra of all three sources are roughly equivalent in the 1.2--1.8~$\micron$ region, but vary in $K$-band peak brightness, with WISE~J2018$-$7423 being the intermediate source.  More importantly, WISE~J2018$-$7423 does not have the broadened 1.05~$\micron$ peak seen in the spectrum of SDSS~J1416+1348B.
This comparison suggests that WISE~J2018$-$7423 is a roughly solar-metallicity field brown dwarf that is  
both older and more massive than the average local population.

\section{Summary}

We have identified five new late-type T dwarfs with WISE,
confirmed through low-resolution, near-infrared spectroscopy with the Magellan FIRE spectrograph.
The spectra indicate classifications ranging from T5.5 to T8.5:, with the latest-type source, WISE~J1812+2721, found to be an excellent match to the T8.5 companion brown dwarf Wolf~940B.
Estimated distances are roughly 12--13~pc, assuming single sources.  Preliminary spectral model fits
indicate {\teff}s as low as 600~K, with a broad range of surface gravities, masses, ages and cloud properties.
In particular, WISE~J1617+1807 and WISE~J2313$-$8037 show indications of being young, low-mass and cloudy based on the relative strengths of their $JHK$ flux peaks, characteristics similar to the 150--800~Myr T8 companion Ross~458C; while the relatively blue and high proper motion T dwarf WISE~J2018$-$7423 may be a solar-metallicity, older and more massive brown dwarf.  Validation of the atmospheric and physical properties of these objects requires more comprehensive broad-band modeling with improved treatment of molecular opacities and the role of clouds.
Nevertheless, it is clear from these early results that WISE will produce an extensive and diverse sample of cool brown dwarfs that can be used to improve our physical understanding of low-temperature, substellar atmospheres (J.\ D.\ Kirkpatrick et al.\ 2011, in preparation).

\acknowledgments

The authors would like to thank telescope operators 
Mauricio Martinez, Sergio Vara, and Jorge Araya at Magellan
for their assistance with the FIRE and LDSS-3 observations, and
T.\ Jarrett from providing scripts and guidance
for the WIRC imaging data reduction.
AJB acknowledges financial support from the Chris and Warren Hellman Fellowship Program.
This publication makes use of data products from the Wide-field Infrared Survey Explorer, which is a joint project of the University of California, Los Angeles, and the Jet Propulsion Laboratory/California Institute of Technology, funded by the National Aeronautics and Space Administration.
This publication also makes use of data products from NEOWISE, which is a project of the Jet Propulsion Laboratory/California Institute of Technology, funded by the Planetary Science Division of the National Aeronautics and Space Administration. 
This publication makes use of data 
from the Two Micron All Sky Survey, which is a
joint project of the University of Massachusetts and the Infrared
Processing and Analysis Center, and funded by the National
Aeronautics and Space Administration and the National Science Foundation.
2MASS data were obtained from the NASA/IPAC Infrared
Science Archive, which is operated by the Jet Propulsion
Laboratory, California Institute of Technology, under contract
with the National Aeronautics and Space Administration.
The Digitized Sky Surveys were produced at the Space Telescope Science Institute under U.S. Government grant NAG W-2166. The images of these surveys are based on photographic data obtained using the Oschin Schmidt Telescope on Palomar Mountain and the UK Schmidt Telescope. The Second Palomar Observatory Sky Survey (POSS-II) was made by the California Institute of Technology with funds from the National Science Foundation, the National Geographic Society, the Sloan Foundation, the Samuel Oschin Foundation, and the Eastman Kodak Corporation. The Oschin Schmidt Telescope is operated by the California Institute of Technology and Palomar Observatory.
This research has also made use of the SIMBAD database,
operated at CDS, Strasbourg, France; the M, L, and T dwarf compendium housed at DwarfArchives.org and maintained by Chris Gelino, Davy Kirkpatrick, and Adam Burgasser; and the SpeX Prism Spectral Libraries, maintained by Adam Burgasser at \url{http://www.browndwarfs.org/spexprism}. 

Facilities: 
\facility{Anglo-Australian Telescope (IRIS2)},
\facility{Fan Mountain (FANCAM)},
\facility{Magellan: Baade (FIRE)},
\facility{Magellan: Clay (LDSS-3)},
\facility{Palomar: Hale (WIRC)},
\facility{SOAR (SpartanIRC)}


\end{document}